\documentclass[structabstract]{aa}

\usepackage{txfonts} 
\usepackage{natbib}
\bibpunct{(}{)}{;}{a}{}{,} 

\usepackage{graphicx}
\usepackage[hang]{subfigure}

\newcommand \sw{{\it Swift}}
\newcommand \ba{{ BATSE}}
\newcommand \ep{$E_{\rm peak}$}
\newcommand \epo{$E^{\rm obs}_{\rm peak}$}
\newcommand \epop{$E^{\rm obs}_{\rm peak}$--P}

\newcommand \gsim{ \lower .75ex \hbox{$\sim$} \llap{\raise .27ex \hbox{$>$}} } 
\newcommand \lsim{ \lower .75ex\hbox{$\sim$} \llap{\raise .27ex \hbox{$<$}} }

\begin{document}
\title{Time resolved spectral behavior of bright BATSE precursors}

\author{D. Burlon \inst{1}\thanks{burlon$@$mpe.mpg.de}
 \and G. Ghirlanda \inst{2} \and G. Ghisellini \inst{2} \and 
J. Greiner \inst{1} \and A. Celotti \inst{3}}
\institute{
Max-Planck-Institut f\"ur Extraterrestrische Physik, 
Giessenbachstra\ss{}e 1, 85740 Garching, Germany 
\and Osservatorio Astronomico di Brera, via E. Bianchi 46, I--23807 Merate, Italy 
\and SISSA, via Beirut 2-4, I-34151 Trieste, Italy}
\date{Received .. ... .. / Accepted .. ... ..} 

\abstract {} {Gamma Ray Bursts (GRBs) are sometimes preceded by 
dimmer emission episodes, called ``precursors", whose nature is 
still a puzzle: they could either have the same origin as the main 
emission episode or they could be due to another mechanism.  
We investigate if precursors have some spectral distinctive feature 
with respect to the main GRB episodes.
}
{To this aim we compare the spectral evolution of the precursor 
with that of the main GRB event. 
We also study if and how the 
spectral parameters, and in particular the peak of the $\nu F_{\nu}$ 
spectrum of time resolved spectra, correlates with the flux.
This allows us to test if the spectra of the precursor and of 
the main event belong to the same correlation (if any).
We searched GRBs with precursor activity in the complete sample of 
2704 bursts detected by \ba\ finding that 12\% of GRBs have one 
or more precursors. 
Among these we considered the bursts with 
time resolved spectral analysis performed by Kaneko et al. 2006, selecting those having at least 
two time resolved spectra for the precursor.} 
{We find that precursors and main events have very similar 
spectral properties. 
The spectral evolution within precursors 
has similar trends as the spectral evolution observed in the 
subsequent peaks. 
Also the typical spectral parameters of the 
precursors are similar to those of the main GRB events. 
Moreover, in several cases we find that within the precursors 
the peak energy of the spectrum is correlated with the flux 
similarly to what happens in the main GRB event. 
This strongly favors models in which the precursor is due to 
the same fireball physics of the main emission episodes. 
} {} 

\maketitle

\section{Introduction}
How does a GRB behave before the onset of the main emission is a debated issue. 
The so--called ``precursor'' activity has been observationally 
addressed by e.g. \citet{koshut95} [hereafter K95], 
\citet{lazzati05} [L05] and \citet{burlon} [B08].

K95 searched in the BATSE sample for precursors defined as 
pulses with a peak intensity lower than that of the main GRB 
and separated from it by a quiescent phase at least as long 
as the duration of the main event. They found precursors in 
$\sim$3\% out of a sample of GRBs detected by \ba\ up to 1994 May. 
Their duration appeared weakly correlated with those of the 
main GRBs and on average shorter than that of the burst. 
The spectral properties of these precursors showed no systematic 
difference  with respect to those of the main GRB event, being both 
softer and harder. However, the comparison of the spectral 
properties of the precursors and of the main event were based on 
the hardness ratio which is only a proxy of the real shape of burst spectra. 

L05 searched for precursors as weak events \emph{preceding} 
the \ba\ trigger. He found, within a sample of 133 bright GRBs, 
that $\sim$20\% showed precursor activity. 
These precursors were on average extremely dimmer than the 
main GRB event, and their durations are weakly correlated 
with that of the main event. In contrast with the results 
of K95, the precursors studied by L05 were softer than the main event.  
Also in this analysis, however, the spectral characterization 
of the precursors were based on the fluence hardness ratio. 
However, given the typically extreme low fluence of most of 
the precursors found by L05, a better spectral characterization 
(e.g. through model fits of a high resolution \ba\ spectrum) 
was almost impossible. 
A difference is how the precursor--to--burst 
separation is measured: K95 consider the time difference between 
the peak of the precursor and that of the main event, while L05 
measure the precursor--to--main event separation from the end of 
the precursor to the start of the GRB.

B08 searched for precursors in the sample of 105 \sw\ GRBs 
with measured redshifts. 
In $\sim$15\% of the sample a precursor was found. 
The definition of precursors adopted in B08 is similar to 
that used by K95. 
The main difference, however, is that B08 
did not require that the precursor precedes the main event by 
an amount of time comparable to the duration of the main event. 
The novelty of B08 was to search and 
study precursors found in a sample of bursts with known redshifts. 
This allowed, for the first time, to characterize the precursor 
energetics and to study how they compare with the main event 
energetics, also as a function of the rest--frame time separation 
between the precursors and the main events.  
The results of B08 suggest that  precursors' spectra 
are consistent with those of the main event. 
Moreover,
regardless of the rest frame duration of the quiescence 
(i.e. the time interval separating the precursor and the burst), 
precursors carry a significant fraction of the total energy ($\approx$30\%) 
of the main event  (see Fig. 1 therein). 
The conclusions of B08 point to a common origin for both precursor and main event. 
Namely, they are nothing but two episodes of the same emission process.

Theoretical models for precursors can be separated into three classes:
the ``fireball precursor'' models \citep{li07, lyutikov03, 
meszaros00, daigne02, ruffini01}; the ``progenitor precursor''
models \citep{ramirezruiz02,lazzati05b} and the ``two step engine'' 
model (\citealp{wang07} [W07], \citealp{lipunova09} [L09]). In the first 
class the precursor is associated to the initially trapped fireball 
radiation being released when transparency is reached.  In the second 
class, based on the collapsar scenario, the precursor is identified 
with the interaction of a weakly relativistic
jet with the stellar envelope. A strong terminal shock breaking out 
of the envelope is expected to produce transient emission. In both 
classes of models the precursors emission is predicted to be thermal, 
characterized by a black--body spectrum.  As for the third class  
in W07 the progenitor collapse leads to the formation of a neutron 
star whose emission would be responsible for the precursor, while 
the star shrinks; subsequent accretion onto the neutron star causes 
its collapse onto a
black hole, originating the GRB prompt. Conversely, in L09 the 
precursor is produced when a collapsing ``spinar'' halts at the 
centrifugal barrier, whereas the main emission is due to a spin--down 
mechanism. Thus, in L09 accretion is not invoked in either steps.

One of the main limitations of K95 and L05 analyses is the poor spectral 
characterization of precursors. They used the hardness ratio HR, 
i.e. the ratio of the counts (or fluences reported in the \ba\ catalogue) 
measured over broad energy channels. However, it is well established 
that the broad band spectra of GRBs can be fitted  by empirical models 
(e.g. Band et al.  1993) composed by low and high spectral power--laws 
with different slopes. The HR is only a proxy of the real spectral 
properties of GRB spectra (e.g. \citealp{ghirla09}), in particular for GRBs with vastly different \ep.  
The other main limitation of these studies, based on the \ba\ GRB catalogue, 
is the lack of redshifts. Indeed, this motivated the study of  B08 of  
\sw\ GRBs with precursors of known redshifts. Nonetheless, 
the spectral analysis of B08 of \sw\--BAT spectra was limited 
by the narrow spectral range (15--150keV): most \sw\ spectra of 
precursors could be fitted by a single power--law (i.e. the peak 
energy of the $\nu F_{\nu}$ spectrum is unknown) and in all 
cases no time resolved spectral analysis of the precursor could be performed. 

The latter point is particularly important: the information 
carried by the strong spectral evolution of GRB spectra 
(e.g. \citealp{ryde05}; \citealp{ghirla02}, \citealp{kaneko06} [K06])  
is completely averaged out when time integrated spectra are 
considered (integrated over the duration of the burst or over the 
duration of single emission episodes, like the precursor and the 
main event in B08). 
An interesting feature found by time resolved analysis of GRB spectra 
is that there could be a positive trend between the spectral peak 
energy \ep\ and the flux $P$ within single emission episodes of GRBs 
\citep{liang04} [L04]. Interestingly, this trend appears similar 
\citep{firmani09} [F09] to that found between the rest frame GRB peak 
energies and their isotropic equivalent luminosities, when considering 
different GRBs with measured $z$  (i.e. so called ``Yonetoku" 
correlation; \citealp{yonetoku04}).  

For these reasons we consider, in this paper, a still unanswered 
question: how does the spectrum of the precursor evolve and how 
does it compare with the evolution of the associated main event? 
In order to answer this question we compare
the time evolution of the spectral 
parameters of precursors and main events.
We also want 
to test if a  possible correlation between the peak energy and the flux, i.e. \epop\  within the precursors exists. 
If this correlation is due to the physics of the emission process or 
to that of the central engine is still to be understood, but if the 
precursors and the main event do follow a similar correlation, this 
would be another piece of the puzzle (in addition to the results of 
B08) suggesting that precursors are nothing else than the first 
emission episodes of the GRB. To this aims spectral data with high 
time and spectral resolution are necessary. \ba\ provides the best 
data for this purpose. 

The paper is organized as follows: in Sec. 2 we describe the sample 
selection and global properties; in Sec. 3 we present the spectral 
comparison between the precursor and the main event within single 
GRBs and we draw our conclusions in Sec. 4.



\section{The sample} 

The Compton Gamma Ray Observatory satellite ({\it{CGRO}}) 
had on board the  Burst Alert and Transient Source Experiment 
(\ba, \citealp{fishman89}), which provided the largest sample of 
GRBs, detected during the 9 yr lifetime. By applying different 
precursor definitions, K95 and L05 searched for \ba\ bursts showing 
a precursor activity. A common feature of these studies is that a 
precursor is a peak separated (i.e. preceding) by a time interval 
and with a lower count rate with respect to the main GRB event. 

The definition of a precursor is somewhat subjective and can easily 
bias the sample. 
L05, by excluding precursors that triggered \ba\, 
selected the faintest precursors. 
K95 instead is missing precursors 
which can be closer than the duration of the rest of the burst.  
For these reasons, consistently with the definition adopted in B08, 
we adopted a definition of ``precursor'' as any peak with a peak flux 
smaller than the main prompt that follows it and that is separated from 
the main event by a quiescent period (namely, a time interval during 
which the background subtracted light curve 
is consistent with zero). We didn't assume {\textit{a priori}} that precursors can occur only in long GRBs (i.e. duration of the main emission episode be $>2$ sec in the observer's frame), albeit in B08 we found no short burst with a precursor.
We adopted 
this ``loose'' definition in order to check, a posteriori, if 
distinguishing characteristics emerge in the analysis. 
This definition is subject to find more easily precursors events 
of the type of K95 in the \ba\ sample.
Since K95 limited the search to half of the \ba\ sample (considering events 
between 910405 and 940529) and due to the slightly different 
precursor definition, we searched for precursors in the complete \ba\ sample.

\begin{figure} 
\resizebox{\hsize}{!}{\includegraphics{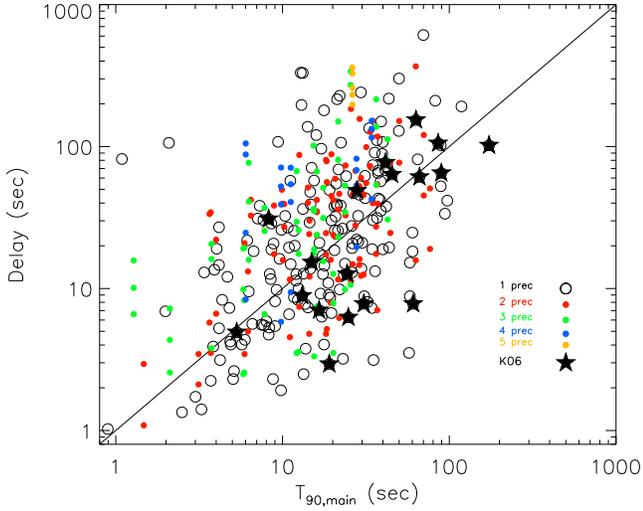}} 
\caption{
Delay (precursors to main event) vs T90 of the main 
prompt emission
for the 264 GRBs with precursors found in the \ba\ sample. Black 
empty circles are GRBs with single precursors (191 cases), 
while filled dots show GRBs with multiple precursors (color 
code as in the legend). Black filled stars represent the 18 
precursors with at least two spectra in K06. The solid line 
represents equality. } 
\label{del_dt} 
\end{figure}

The final BATSE GRB 
sample\footnote{http://heasarc.gsfc.nasa.gov/docs/cgro/batse/BATSE\_Ctlg/basic.html} 
contains 2704 GRBs. We found 2121 GRBs out of 2704 total triggers for which 
there was a 64 ms binned light 
curve\footnote{http://[...]/batse/batseburst/sixtyfour\_ms/index.html} available. 
We inspected the background subtracted light curve of each GRB and 
found 264 GRBs (12.5\%) with a precursor. The majority (191) of GRBs 
showed one precursor, 48 showed double precursors, 19 showed three 
precursors, 5 showed four precursors and in only one case we found 
five precursors, according to our definition. 

\subsection{Sample properties}

From the 64 ms \ba\ light curves we calculated the duration of the  
precursor and main emission event for each of the 264 GRBs with precursors. 
The duration was defined as in the \ba\ GRB catalogue, i.e. T90. 
This corresponds to an integral measure, being the time interval containing 
the 90\% (from 5\% to 95\%) of the counts inside each peak considered, either precursor or main event. 
  
We define the time delay between the precursor and the main event as
the difference between the beginning of the main event and the end
time of the precursor. The mean durations of
    precursors and main emission episodes are $\sim$15 s and $\sim$24
    s respectively. The mean duration of the delays is $\sim$50 s.

\begin{figure} 
\resizebox{\hsize}{!}{\includegraphics{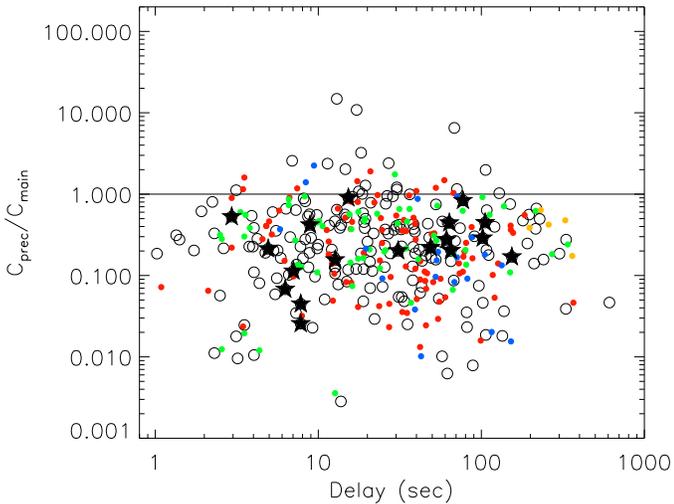}} 
\caption{Ratio of 
precursor to main event counts
versus  delay times. 
Symbols and colour codes are as in Fig. \ref{del_dt}.} 
\label{del_rat} 
\end{figure} 

In Fig. \ref{del_dt} we show the delays of the precursors versus the
duration T90 of the main GRB for the 264 GRB with precursors.  The
probability of a chance correlation among the duration of the
    GRBs with a single precursor (open circles and filled star symbols
    in Fig.\ref{del_dt}) and the corresponding delay is $3.53 \times
    10^{-14}$. An even lower chance probability is found including
    also GRBs with multiple precursors.


Since we do not know the redshift of these GRBs, we cannot exclude that the
correlation is at least in part the result of the common redshift
dependence of both the delay and the T90.
Moreover, Fig. \ref{del_dt} shows no apparent difference between GRBs with
single or multiple precursors.  This result is somewhat different from that
reported by \citet{ramirezruiz01a}.  By investigating the temporal
properties of multi--peaked GRBs (but note that they put no particular emphasis on
precursors) they found a strong one--to--one correlation (4$\sigma$
consistency) 
between the duration of a peak and the duration of the quiescence time
interval before it.

In Fig. \ref{del_rat} we show the ratio of the total counts
(integrated over T90) of each precursor with respect to the counts in
the corresponding main GRB plotted as a function of the delay time. In
most cases the precursor total counts are a fraction (of the order
10--20\%) of the counts of the main GRB.  Also in this case we do not
find any difference between single precursors and multiple ones. Not
surprisingly, a handful of GRBs show a precursor stronger than the main
emission. In these cases, typically the precursor has a duration
  much larger than that of the main which over-compesates its lower
  peak flux, thus giving a higher integral count number for the
  precursor with respect to the main event.

\begin{figure} 
\resizebox{\hsize}{!}{\includegraphics{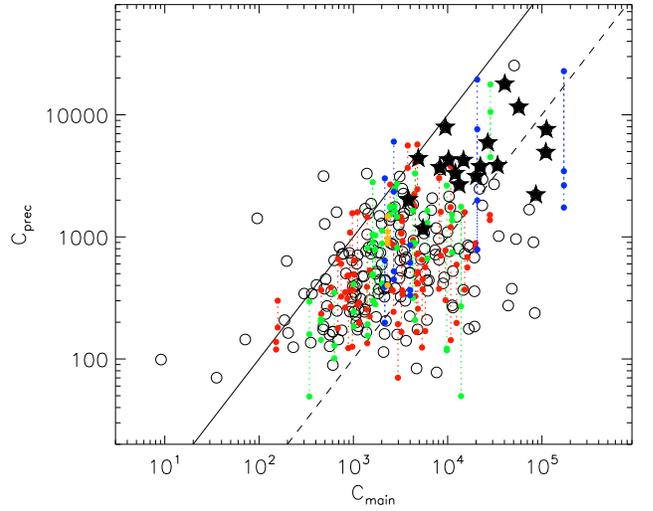}} 
\caption{
Total counts of the precursor versus the total counts
of main event. 
Symbols and colour codes are as in Fig. \ref{del_dt}. 
The solid line represents equality, while the dashed line
corresponds to precursors dimmer than main events by a factor 10.
} 
\label{fluence} 
\end{figure}

Fig. \ref{fluence} shows the total counts of the precursors 
with respect to the total counts in the main GRBs. 
In this plane different selection cuts are evident. 
The selection criterion for defining 
precursors in this work is evident as the lack of precursors to 
the left of the equality line (solid).

 It is apparent from Fig. \ref{del_rat} and Fig. \ref{fluence} that neither the delay times of the precursors with respect to the onset of the main event, nor the integrated counts of the peaks seem to show a specific clustering. Therefore, we can rule out the existence of a sub-class of ``real'' precursors among the complete sample, given the selection method.

\section{Spectral evolution}

In order to study the spectral evolution of the precursors and compare
it with that of the main event, we rely on the time resolved spectral
catalogue of \citet{kaneko06}.  K06 analyzed the spectra of selected
bright \ba\ GRBs.  These were selected to have a peak photon flux (on
the 256 ms time scale and integrated in the 50--300 keV) greater than
10 photons cm$^{-2}$ s$^{-1}$ or a total energy fluence greater than
$2.0\times 10^{-5}$ erg cm$^{-2}$ in the energy range $\sim$20--2000 keV.  This
mixed criterion ensured K06 to have a minimum number of time
  resolved spectra distributed within the duration of each GRB so to
  study the features of its spectral evolution with sufficient
  details.
This led to a sample of 350 GRB. 

For most GRBs the high energy resolution data of the LAD detectors were analyzed. 
These data consists of $\sim$ 128 energy channels distributed between 
$\sim 30$ keV  and 2 MeV accumulated during the burst with a minimum 
time resolution of 128 ms. 
In some cases also lower energy resolution data (MER) were analyzed. 
K06 fitted both the time integrated spectra 
and the time resolved spectra with 5 
different spectral models: a 
simple power--law (PWR), the Band model \citep{band93} (BAND), a Band model with fixed high energy power law component $\beta$ (BETA), a power--law 
with an exponential cutoff at high energies (COMP), or a smoothly broken 
power--law (SBPL). The spectra within a single GRB were accumulated in 
time according to a minimum S/N ratio (required to be larger than 45 in 
each time resolved spectrum, integrated over the energy range 30 and 2000 keV). 
In the final catalogue of K06 the best fit parameters for all the fitted 
models are given for all the time resolved spectra within a single burst. 
Through this large data set, it is possible to construct the time evolution 
of the spectral parameters of the bursts.  

We cross--checked the sample of K06 with the 264 GRB with precursors 
that we have found in the \ba\ catalogue. 
We found 51 GRBs with precursors 
with time resolved analysis reported in the K06 sample. 
However, 
since
our aim is to characterize how the 
spectrum of the precursor evolves in time, we restricted this sample 
to those GRBs with at least 2 time resolved spectra analyzed by K06 
in the time interval of the precursor. 
This condition reduces the sample to 18 GRBs. 
All these have a single precursor in their light curve (except for 
trigger \#6472, that has two precursors). 
In Figs. \ref{del_dt}, \ref{del_rat}, and \ref{fluence} 
these 18 events 
are shown (star symbols):
they correspond to the 
bright end of the distribution of count fluence of the precursors.

%
%
%

 \begin{figure}
 \centering
 \subfigure[The top panel shows the complete light--curve in units of 
count--rate and just below is the same light curve in physical units 
(erg cm$^{-2}$s$^{-1}$) binned into time intervals corresponding to 
the time resolved spectra extracted and analyzed by K06. 
The mid panels show the evolution of the spectral parameters 
of the COMP model, i.e. the low energy photon spectral index 
$\alpha$ and the peak energy of the $\nu{\rm F}\nu$ spectrum  
(\ep\ ). These correspond to the zoom in the time interval of the 
precursor and of the main event (colour symbols corresponding to 
the precursor).]
{\includegraphics[width=9cm,height=9.4cm]{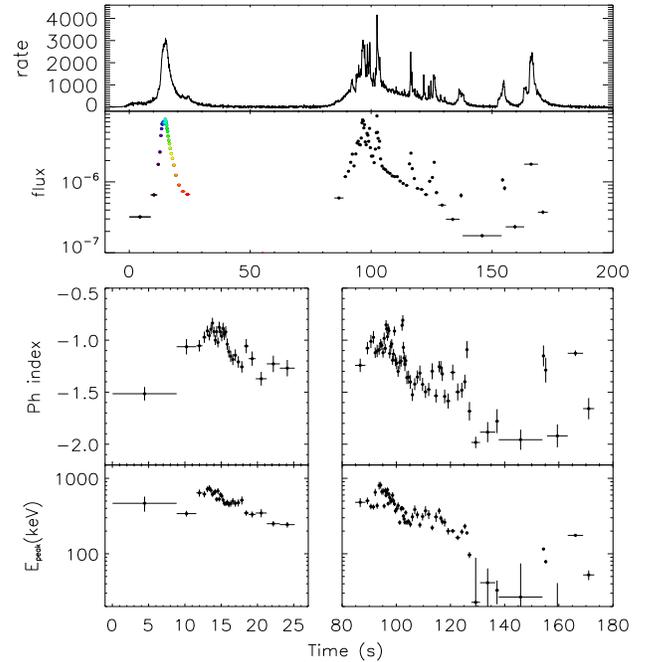}}
 \subfigure[Panels show $\alpha$ ({\it{top}}) and \epo\ ({\it{bottom}}) versus the flux. 
The spectral parameters of the precursor are shown with filled stars and 
joined by a dashed line. The first spectrum is the black one. 
The spectral parameters of the main emission episode are shown 
with empty circles.]
{\includegraphics[width=9cm,height=9.4cm]{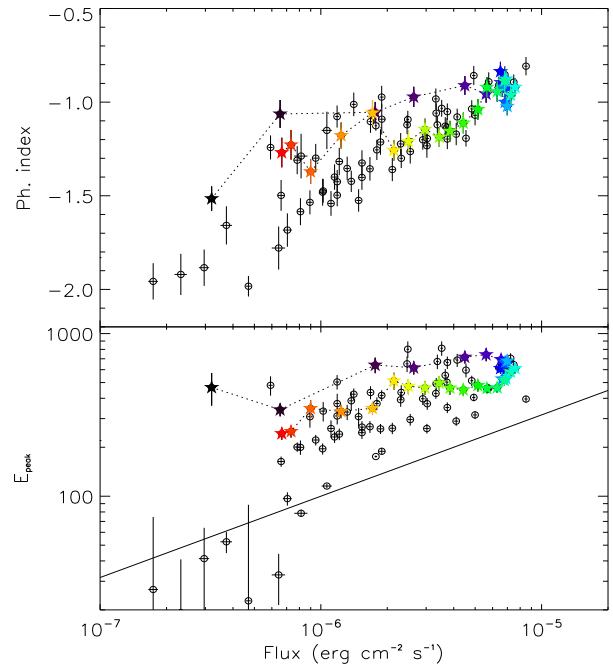}}
 \caption{GRB 930201 (trigger \#2156)  }
 \label{2156}
 \end{figure}
 
 For these 18 GRBs with time resolved spectral analysis reported in
 K06 we show (panel (a) of Fig. \ref{2156}, Fig. \ref{7688fg1} and
 following even figures) the light curve in counts (and in physical
 units as obtained by the spectral analysis) and the time evolution of
 the best fit parameters. It has been shown that when analyzing time
 resolved \ba\ spectra, especially for S/N $\lsim$ 80 (e.g. K06), the
 best fit model is often a cutoff power--law.  This might be due to
 the difficulty of constraining the best fit parameter of the BAND
 model (i.e. the high energy spectral index of the power--law) when
 the fluence of the spectrum is low (as systematically expected in
 time resolved spectra with respect to time integrated ones).  For
 this reason we decided to plot for all the 18 GRBs the spectral
 results given by K06 of the fit with the COMP model.  In some cases
this is not the best fit model of the time resolved spectra but for the
 aims of the present analysis, i.e.  the {\emph{relative}} comparison of the
 spectral evolution of precursors with respect to that of main bursts,
 any systematic effect due to the fit of the spectra with the COMP
 model is not affecting our conclusions.  We show in Fig. \ref{2156}
 that both the photon spectral index and \epo\ follow a strong
 soft--to--hard evolution in the rising part of the precursor, and
 vice-versa in the descending part.  In the main emission event both
 spectral parameters show a general hard--to--soft trend, but inside
 each peak they \emph{both} follow the same trend shown inside the
 single peak of the precursor and moreover they track the flux.

The latter consideration is shown in panel (b) of Fig. \ref{2156} 
(see for comparison the lower panel of Fig. \ref{7688fg2} and following odd 
figures) where a correlation between the peak energy \epo\ and the 
flux P is apparent.  Note however that GRB 930201 is the case with best statistics and hence does not necessarily stand for a general behavior. 
We connected (dashed line) the evolution of the 
spectral parameters only inside the precursor. 
The colour code is as 
in panel (a): namely, the first (last) spectrum is the black (red) one.
%
%
It has been recently pointed out (e.g. \citealp{borgonovo01},
\citealp{liang04}, and more recently by \citealp{firmani09} for \sw\ GRBs),
that when considering the spectral evolution of long GRBs there is a
trend between the evolution of the flux $P$ and the peak energy \epo\
i.e.  approximately $P \propto {\rm E}_{\rm{peak, obs}}^{\gamma}$.  In
particular Firmani et al. (2009) show that 84\% of the K06 sample have
$\gamma \sim 2$ at the 3$\sigma$ level. In addition, the correlation
is not biased systematically by the value of $P$, though its
uncertainty increases with decreasing flux.  We can fiducially
extrapolate this evidence to precursors, keeping open the question of
identifying the hidden physical mechanism that determines the value of
$\gamma$.

Intriguingly, this is similar to the correlation between the peak 
luminosity and the peak energy (time integrated over the duration of the burst) 
in GRBs with measured redshifts (so called ``Yonetoku'' correlation). 
A similar result was reached by Liang et al. (2004) based on the 
spectral evolution of the brightest \ba\ GRBs but for which no 
redshift was measured.  Again, when studying the correlation 
between the luminosity and the peak energy within the few GRBs 
detected by \ba\ and with known $z$, \citet{firmani09}  
finds that the correlation is present. The existence of a correlation 
within a single GRB similar to the Yonetoku correlation could be 
indicative of a physical origin for the quadratic link between the 
flux and the peak energy.

We can test if and how such a correlation holds in the GRBs with precursors 
that we have considered and/or if the \epo\ and $P$ of the precursor are consistent 
with the correlation defined by the prompt. 

If this correlation is due to the physics of the emission process or 
to that of the central engine is still to be understood, but if the 
precursors and the main event do follow a similar correlation, this 
would be another piece of the puzzle suggesting that precursors are 
nothing else than the first emission episodes of the GRB.


\section{Discussion}

\begin{figure}
\resizebox{\hsize}{!}{\includegraphics[]{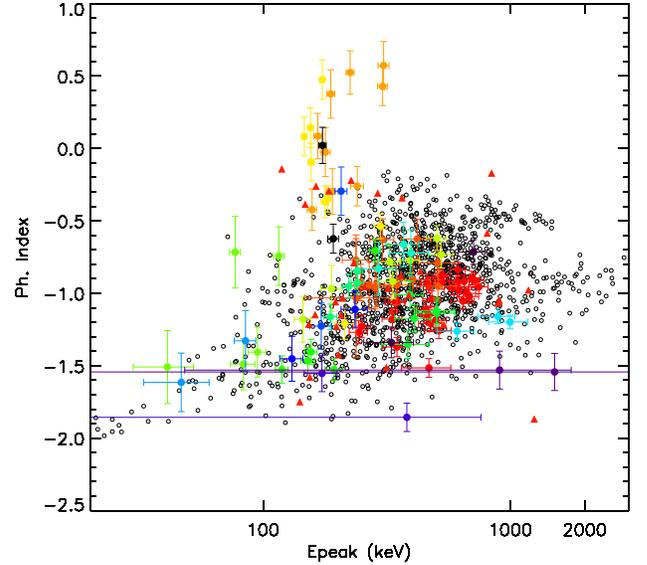} }
\caption{
Photon spectral indices $\alpha$ versus the peak energy  \epo\
for 51 GRBs with precursor. 
Spectral parameters correspond to the time bins of the
time resolved spectra extracted and analyzed by K06. 
The 18 bursts belonging to our sample are shown with filled dots 
(different colors represent different GRBs). 
We added also 33 precursors with a single spectrum data point 
(red triangles). Spectral parameters of the main emission episode are shown 
in black empty circles.  } 
\label{correla1} 
\end{figure}
%
\begin{figure}
\resizebox{\hsize}{!}{\includegraphics[]{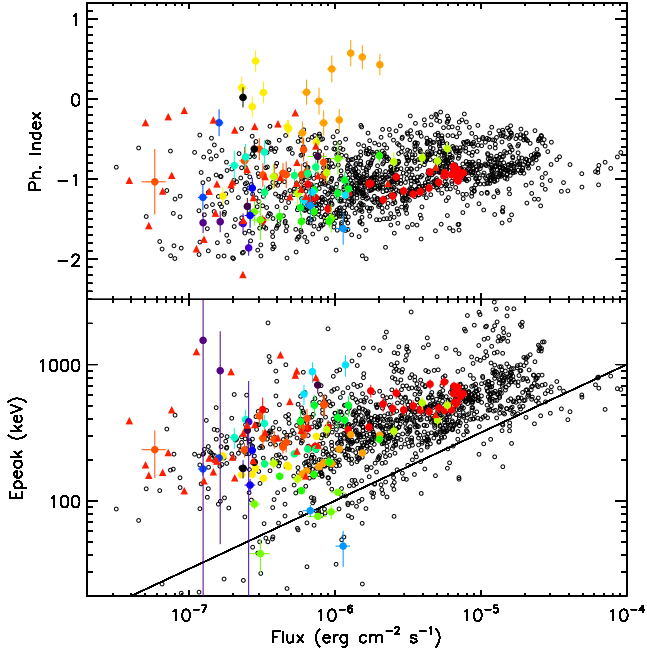} }
\vskip -0.5 cm
\caption{
The photon spectral index $\alpha$ ({\it{top}}) and \epo\ 
({\it{bottom}}) versus the flux $P$ in erg cm$^{-2}$s$^{-1}$. 
The solid line in the bottom panel corresponds to 
$P \propto {\rm E}_{\rm{peak,obs}}^2$. 
Colour code and symbols as in Fig. \ref{correla1}. } 
\label{correla2} 
\end{figure}

Figs. \ref{correla1} 
shows the photon spectral indices $\alpha$ versus the peak energy \epo\
for all 51 GRBs with precursor present in K06,
while Fig. \ref{correla2}
shows for the same bursts how $\alpha$ and \epo\ 
behave with the flux $P$.
Different symbols (and colors, in the electronic edition)
marks the precursor and the main event points.
Filled symbols correspond to the 18 GRBs with at least two
spectra for the precursors.
Red triangles mark the 
remaining precursors in K06 with just one spectrum.
Empty black dots correspond to the spectral parameters of the main events.
 

Fig. \ref{correla1} shows that
on average precursors and main GRB emission 
episodes span the same parameter space, while  Fig. \ref{correla2} shows
that they follow similar correlations with the flux.

\begin{figure}
\resizebox{\hsize}{!}{\includegraphics[]{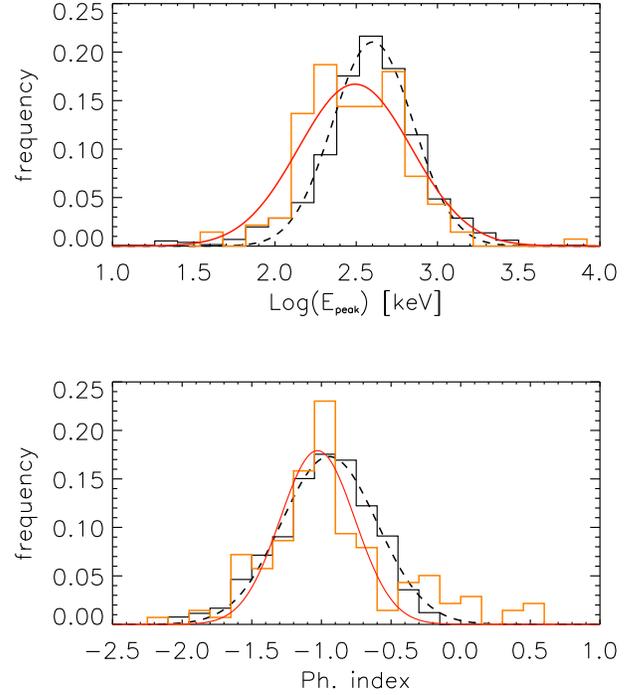} }
\caption{ Normalized distribution of \epo\ ({\it{top}}) and of 
the spectral photon indices $\alpha$ ({\it{bottom}})
of precursors and main emission episodes. Precursor distribution and gaussian fit are 
shown in colored filled line, main emission episode in black dashed line. } 
\label{distrib} 
\end{figure}

\begin{figure} 
\resizebox{\hsize}{!}{\includegraphics{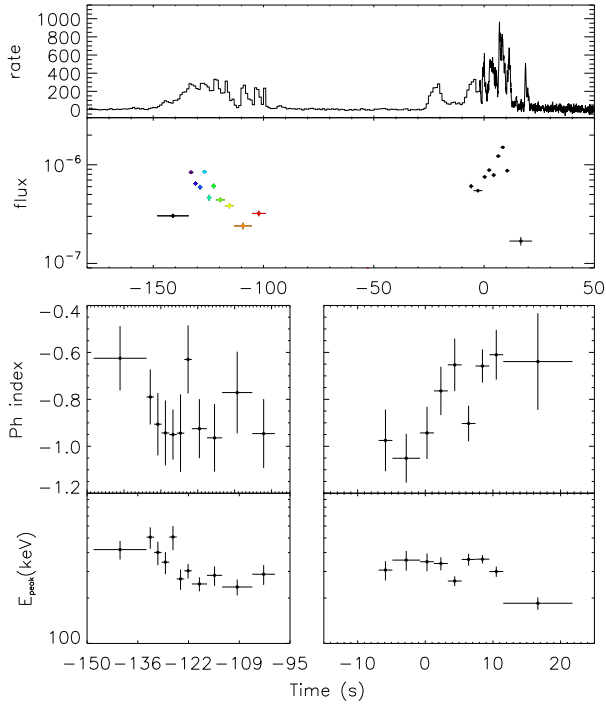}} 
\caption{Trigger \#7688. Colour code and description as in Fig. \ref{2156}--a } 
\label{7688fg1} 
\end{figure}

The distributions of the low energy photon indices $\alpha$ 
of the precursors and the main events are roughly consistent 
(the Kolmogorov-Smirnov KS null hypothesis probability  is $\simeq 10^{-2}$).  
Fitting the two distributions (see Fig. \ref{distrib}, upper panel) with 
gaussian profiles we find      
$\langle \alpha_{\rm prec}\rangle = -1.03\pm  0.27$ and 
$\langle \alpha_{\rm main}\rangle = -0.94 \pm 0.34$.

Three (\#5486, \#6472, \#7343) of the 18 GRBs studied here present extremely 
hard spectra. 
One of them, i.e. GRB 960605 (\#5486, see Fig. \ref{5486fg2}), 
could even be consistent with a black--body spectrum at the 
very beginning of the precursor. 
These few cases populate the  
upper part of Figs. \ref{correla1} and \ref{correla2} (upper panel). 
We have re-extracted the LAD data for this burst and 
reanalyzed them. 
We confirm the findings of K06. 
The finding of a precursor with a spectrum consistent with a black--body
should not be taken as a proof of a radical difference with the
main event, since it has been already
pointed out (e.g., \citealp{ghirla03}) that a non--negligible fraction of 
GRB ($\sim$5\%) start their emission with a black body spectrum.  

Comparing the distributions of $\log$(\epo) we find that they
are somewhat different (K--S null hypothesis probability $\sim 10^{-4}$).
%
Fitting again with gaussian profiles the 
two distributions in Fig. \ref{distrib} (lower panel) we find the mean value and 1$\sigma$ 
scatter for precursors: $\log$(\epo) = 2.49$\pm0.35$ to be compared
to $\log$(\epo) = 2.60$\pm 0.24$ for the main emission events.
The distribution of \epo\ for the precursors is slightly softer 
than the one of the main prompt emission.
This result is not surprising when looking at the bottom panel of Fig. \ref{correla2}: 
the peak energy of precursors seem to follow the trend (when \epo\ is plotted 
with respect to flux) drawn by the GRB main emission, but at the lower left end 
of the track. 

\begin{figure} 
\resizebox{\hsize}{!}{\includegraphics{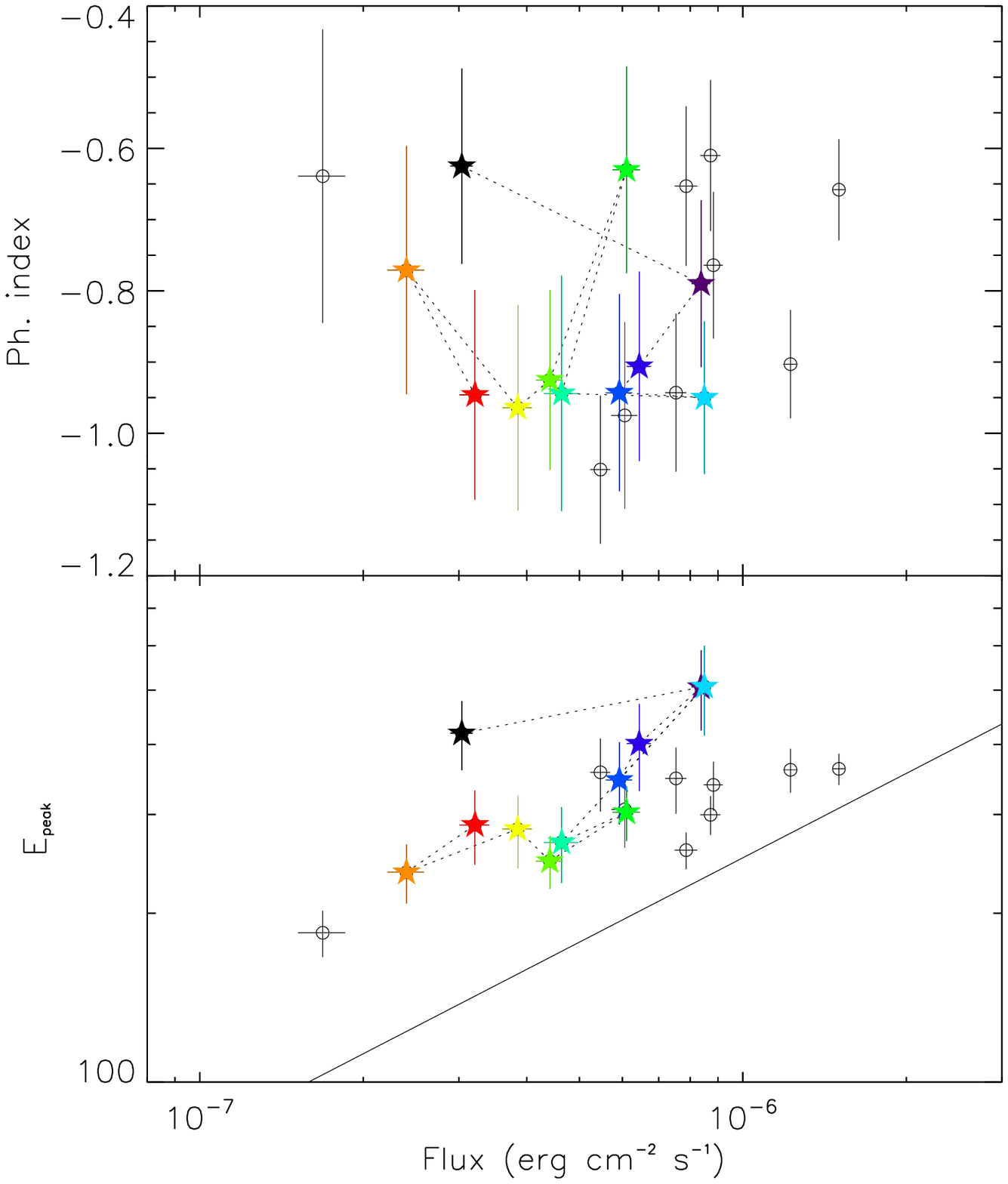}} 
\caption{Trigger \#7688. Colour code and description as in Fig. \ref{2156}--b} 
\label{7688fg2} 
\end{figure}

In the 7 precursors with more time resolved spectra 
(\#2156, \#7688, \#5486, \#6472, \#3481, \#3241, \#1676), 
\epo\ shows a strong evolution but nonetheless is always 
consistent with the correlation drawn by the main event (as shown in 
Fig. \ref{2156}--b (see lower panel in Figs. \ref{7688fg2}, 
\ref{5486fg2}, \ref{6472fg2}, \ref{3481fg2}, \ref{3241fg2}, \ref{1676fg2}). 
Note that these similar trends in the evolution of \epo\ do not 
depend upon the delay, as these vary among $\sim$9 s (for \#1676) 
and $\sim$75 s (for \#7688). 
%
%
Note that at odds with B08, this consideration is based only on observed 
time intervals, because the redshift $z$ is unknown for all GRBs in this work.
Two of them, namely \#2156 and \#1676, also show consistent 
evolution in $\alpha$ between the precursor and the main event 
(see upper panels of Figs. \ref{2156}--b and \ref{1676fg2}). 
The other 5 GRBs (of this group of 7) show an evolution in $\alpha$ which is different
in the precursor and in the main event: in two cases 
(\#5486 and \#6472) $\alpha$ starts extremely hard and evolves to softer 
values (see upper panel of  Figs. \ref{5486fg2} and \ref{6472fg2}). 
In the last three cases (\#3481, \#3241 and \#7688) either the photon 
spectral index evolves in a different way with respect to the one of 
the main emission episode (as in Figs. \ref{3241fg2} and \ref{7688fg2}, 
upper panels), or it lies in a different region of the parameter space 
(see upper panel of Fig. \ref{3481fg2}).

The remaining 11 GRBs of our sample have more coarsely sampled precursor spectra. 
The trend of \epo\ of the precursor 
is consistent with that of the main event in 8 cases. 
In \#3253, \#6454, \#3057, \#4368, \#1157, \#6629, \#3301, \#7343 
(see upper panels of Figs. \ref{3253fg2}, \ref{6454fg2}, \ref{3057fg2},
\ref{4368fg2}, \ref{1157fg2}, \ref{6629fg2}, \ref{3301fg2}, \ref{7343fg2}) 
the peak energy in the spectra of precursors follow the same correlation
with the flux drawn by the main emission. 
Notwithstanding, the number of 
spectra extracted by K06 in the precursor varies between five and two, 
thus preventing any more confident claim. 
Among these 8 GRBs, in the latter 3 the photon spectral indices 
$\alpha$ of the precursors do not 
track the trend drawn by the main emission event (see upper panels of 
Figs. \ref{6629fg2}, \ref{3301fg2}, and \ref{7343fg2}), being always 
softer (with the exception of the onset of the precursor in \#7343, 
which has $\alpha \simeq 0$). 
In the former badly sampled 5 GRBs, 
also $\alpha$ is consistent with the trend described by the spectra 
of the main impulse.  
Note that also in these 8 cases the delay does 
not represent a distinguishing feature, as it can vary from 7 s (e.g. \#3253) 
up to $> 100$ s (\#3663).

The last three GRBs, namely \#3663, \#2700, and \#3448 present 
hardly distinguishable spectral characteristics (i.e., both $\alpha$ and \epo). 
This is due either to the extremely low number of spectra extracted in the 
precursor, or in the main impulse, or both at the same time (see Figs. 
\ref{3663fg2}, \ref{2700fg2}, \ref{3448fg2}). 
In our opinion this prevents any further claim.

\section{Conclusion}

In this work we presented, for the first time, a time resolved 
spectral analysis of bright precursors based on spectral parameters, 
namely the photon spectral indices $\alpha$ and the observed peak energy \epo. 
This was done by using High Energy Resolution spectra extracted by 
K06 in a sample of 350 bright GRBs out of the complete sample of 2704 
confirmed GRBs observed by the  \ba\ instrument. 
Of the 51 bursts with precursor present in K06,
we selected a sample of 18 GRBs having at least two time
resolved spectra of the precursor.


The comparison with the main emission episode has three outcomes. 
The first is that the photon spectral indices of precursors and main events 
are consistent, while the 
peak energies of the precursors
are mildly softer (see Fig. \ref{distrib}). 
Secondly, both $\alpha$ and \epo\ do show an evolution 
(extreme in a handful of cases) that defines 
a relation between the flux $P$ 
and the spectral parameters (note that the $P-E_{\rm{peak,obs}}^{\gamma}$ 
correlation was recently reported (e.g. F09) regardless the presence 
of precursors). 
Finally we showed that delays do not represent a 
distinguishing feature in the trend of $\alpha$ or \epo.

We found one GRB (out of 18) in which the onset of the emission of 
the precursor is consistent with black--body emission 
(i.e., \#5486 -- see Fig. \ref{5486fg2}). 
This was expected, since 
\citet{ghirla03} showed that 5\% of \ba\ GRBs show extremely hard 
emission at the onset of the first impulse.

Moreover, comparing the integrated counts in the peaks of precursors 
with respect to the ones of the main impulses, we confirmed the results 
of B08 (see Fig. \ref{fluence}). 
Indeed precursors carry a significant 
fraction of the energy of the main emission episode, regardless the 
duration of the time interval of quiescence.

These results, in addition to B08, point strongly to the conclusion 
that the onset of emission of GRBs (called precursor), even if separated 
from the main emission episode by hundreds of seconds (in the observers 
frame), is indistinguishable from that of the main event. 
Moreover the delay remains a puzzling issue. 
This suggests that we should reconsider the 
idea of what a precursor is. 
Since our result is partially in contrast with L05 we cannot rule out the possibility 
that ``real precursors'' belong to another class of very dim pulses of 
different origin. 
Nonetheless, both kind of precursors can show very 
long delays, thus tackling any theoretical model for GRB prompt emission.


\begin{acknowledgements} 
We acknowledge Marco Nardini and Lara Nava for stimulating discussion. D.B. is supported through DLR 50 OR 0405. 
This research was partially supported by ASI-INAF I/088/06/0 and MIUR. 
We acknowledge the use of public data from the \ba\ data archive. D.B. thanks the OAB for the kind ospitality during the completion of this work.
\end{acknowledgements}


\clearpage


\begin{figure} 
\resizebox{\hsize}{!}{\includegraphics{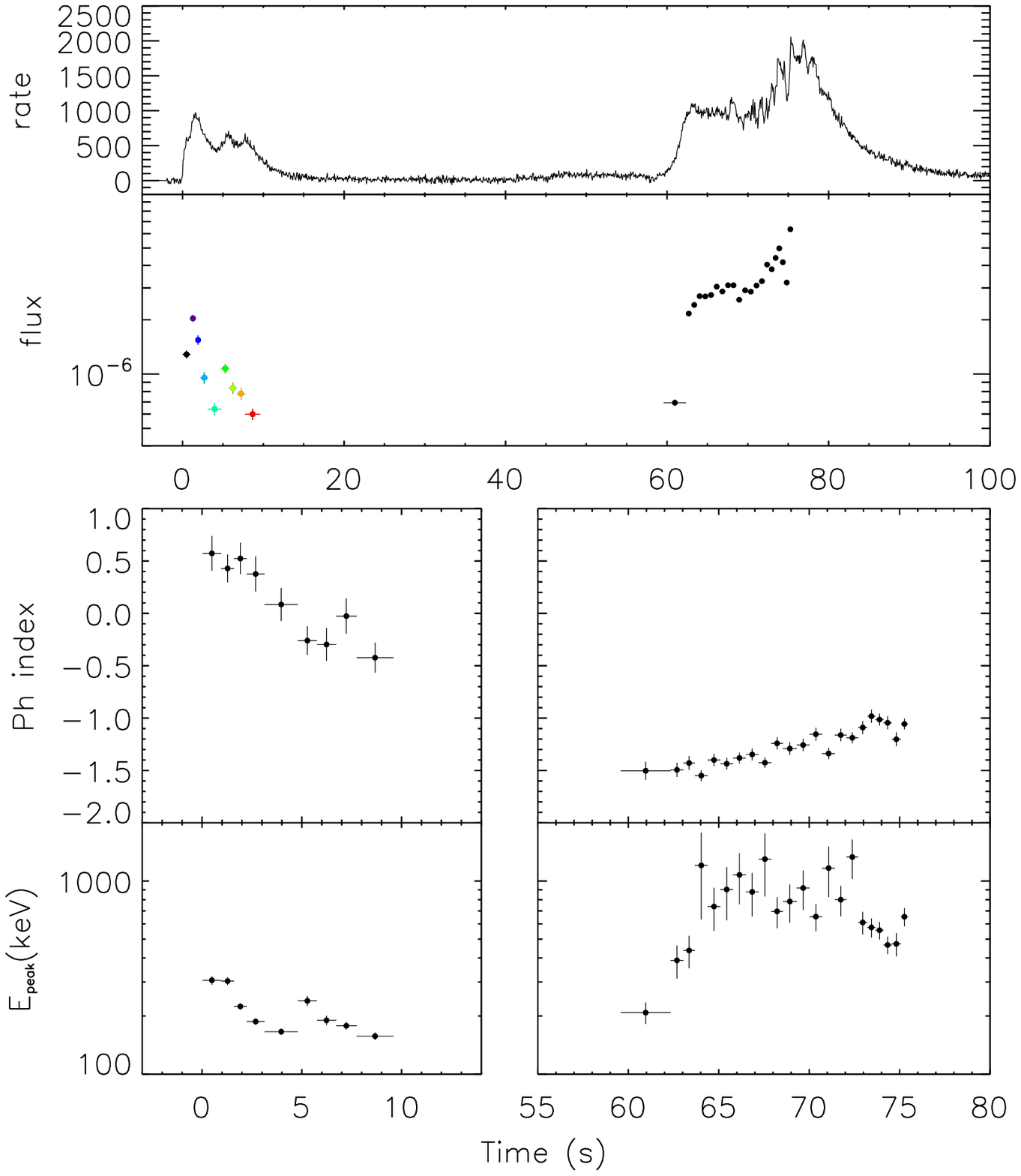}} 
\caption{Trigger \#5486. Colour code and description as in Fig. \ref{2156}--a} 
\label{5486fg1} 
\end{figure} 
\begin{figure} 
\resizebox{\hsize}{!}{\includegraphics{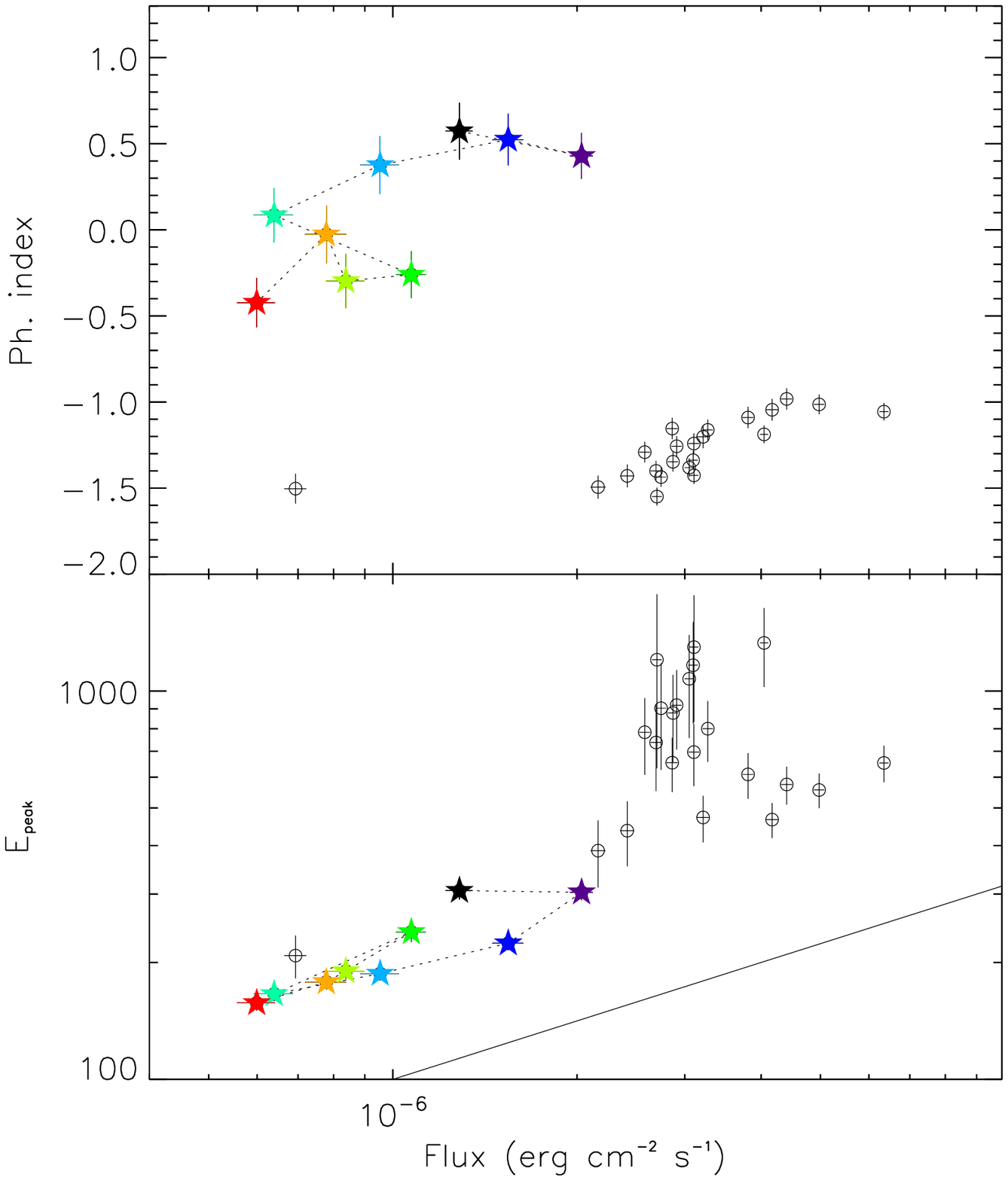}} 
\caption{Trigger \#5486. Colour code and description as in Fig. \ref{2156}--b} 
\label{5486fg2} 
\end{figure}

\begin{figure} 
\resizebox{\hsize}{!}{\includegraphics{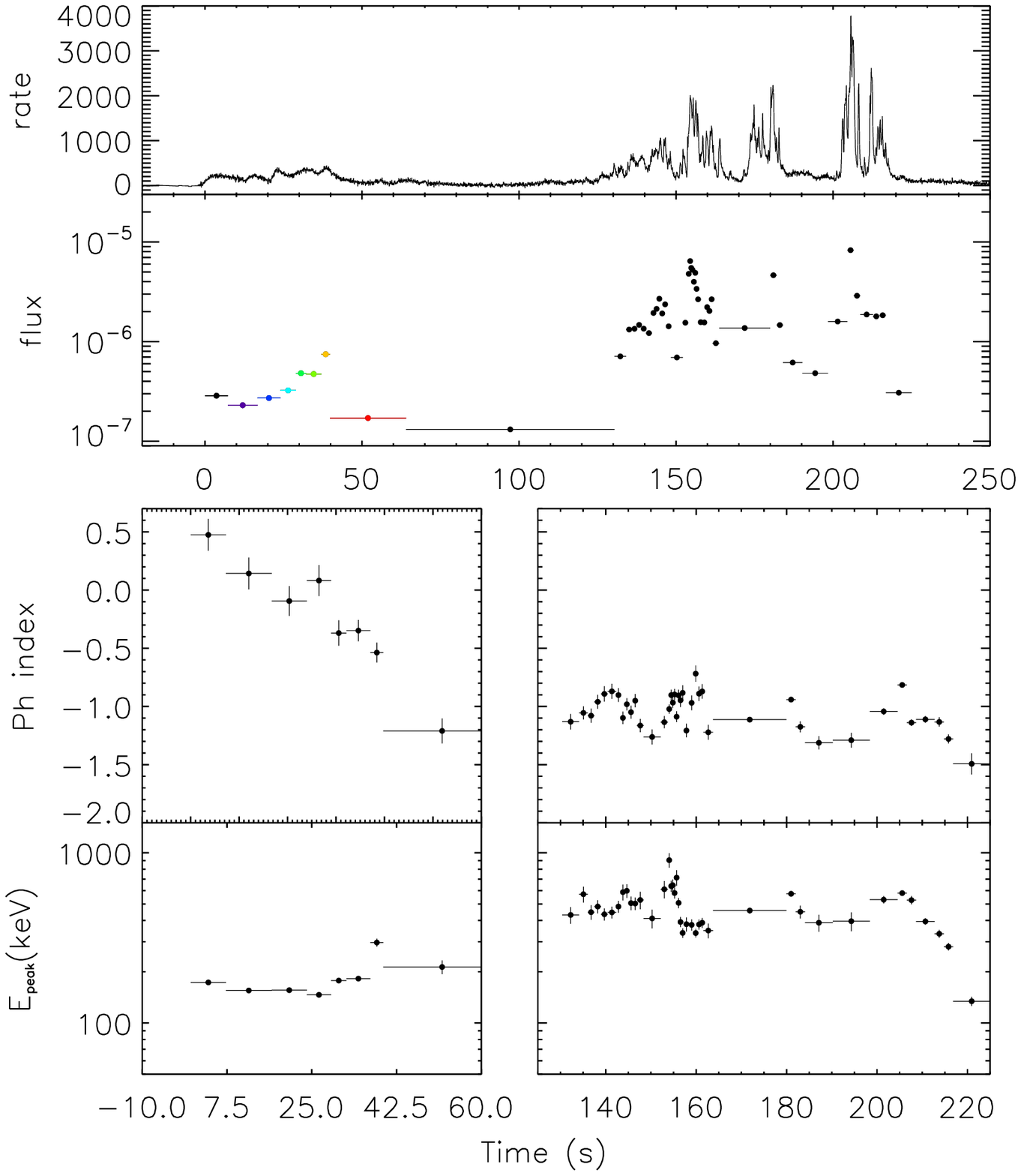}} 
\caption{Trigger \#6472. Colour code and description as in Fig. \ref{2156}--a} 
\label{6472fg1} 
\end{figure} 
\begin{figure} 
\resizebox{\hsize}{!}{\includegraphics{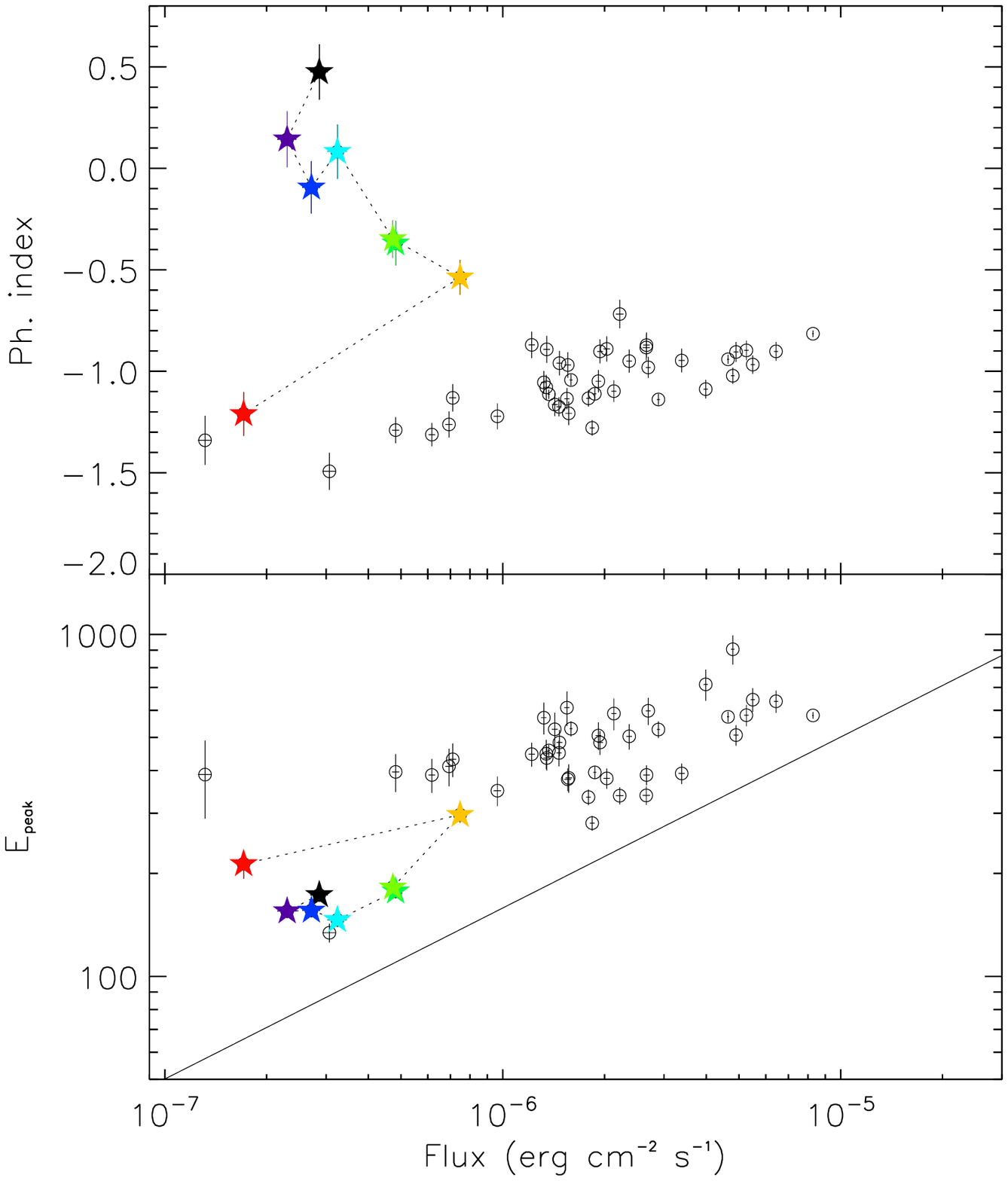}} 
\caption{Trigger \#6472. Colour code and description as in Fig. \ref{2156}--b} 
\label{6472fg2} 
\end{figure}

\begin{figure} 
\resizebox{\hsize}{!}{\includegraphics{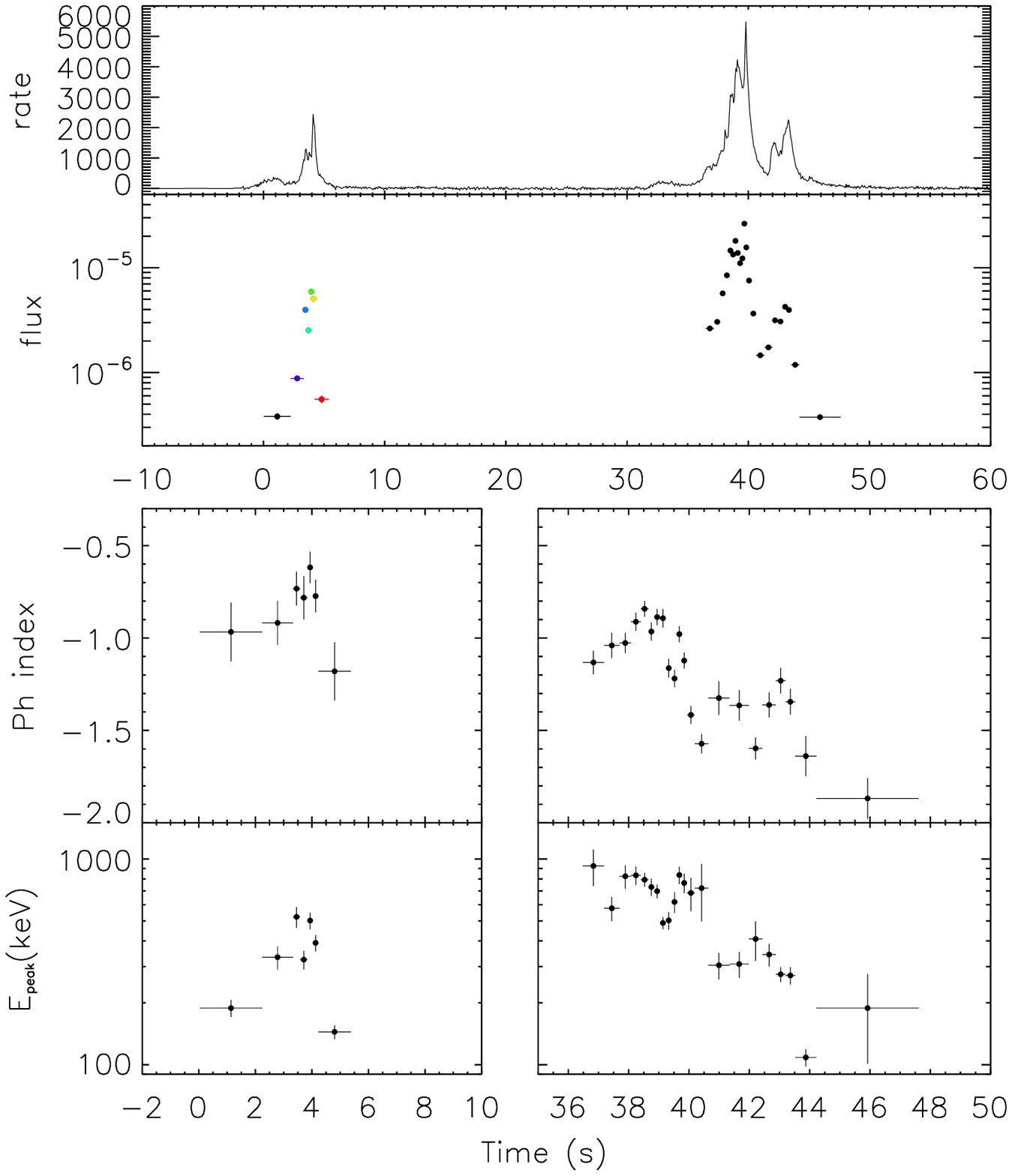}} 
\caption{Trigger \#3481. Colour code and description as in Fig. \ref{2156}--a} 
\label{3481fg1} 
\end{figure} 
\begin{figure} 
\resizebox{\hsize}{!}{\includegraphics{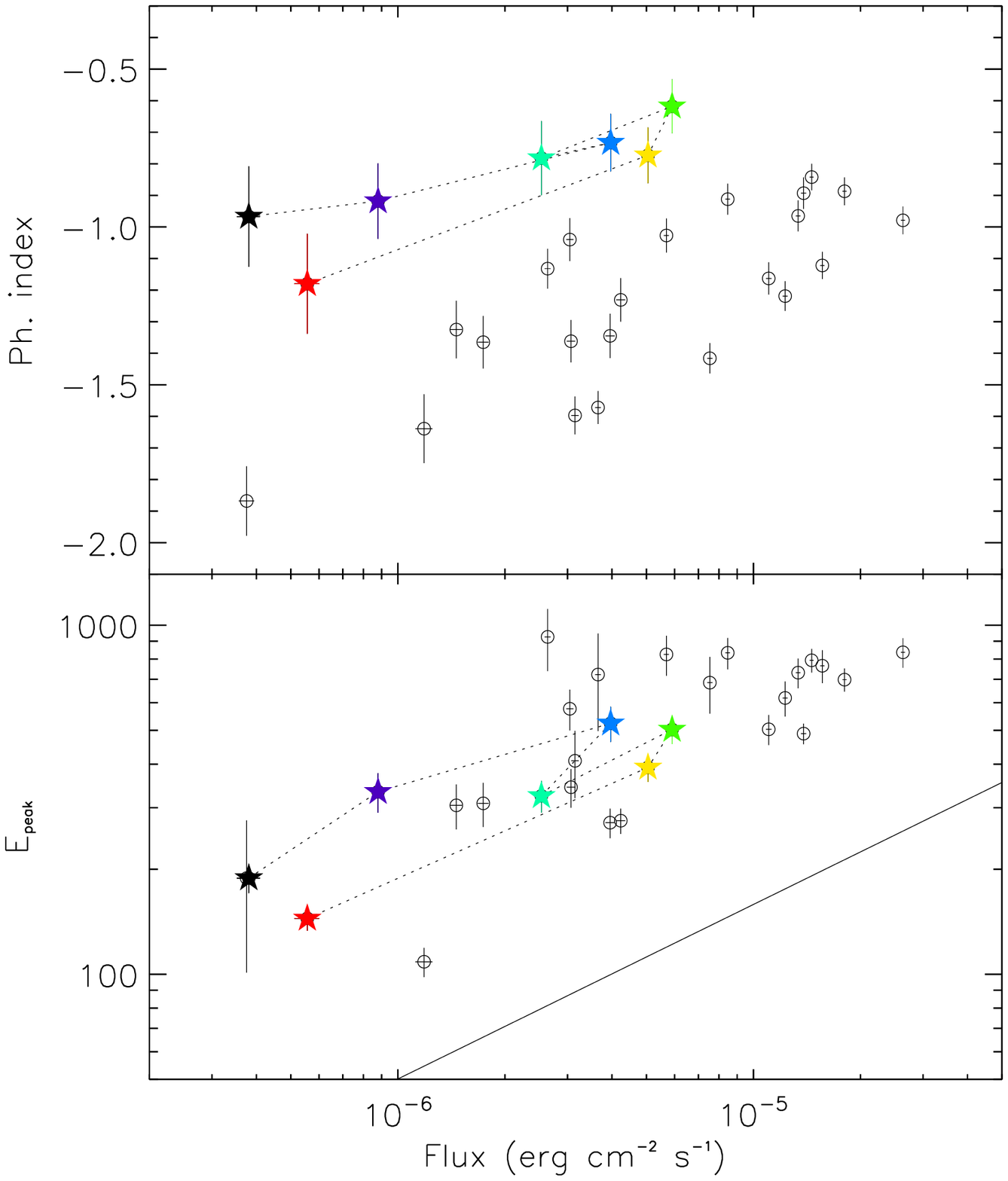}} 
\caption{Trigger \#3481. Colour code and description as in Fig. \ref{2156}--b} 
\label{3481fg2} 
\end{figure}

\begin{figure} 
\resizebox{\hsize}{!}{\includegraphics{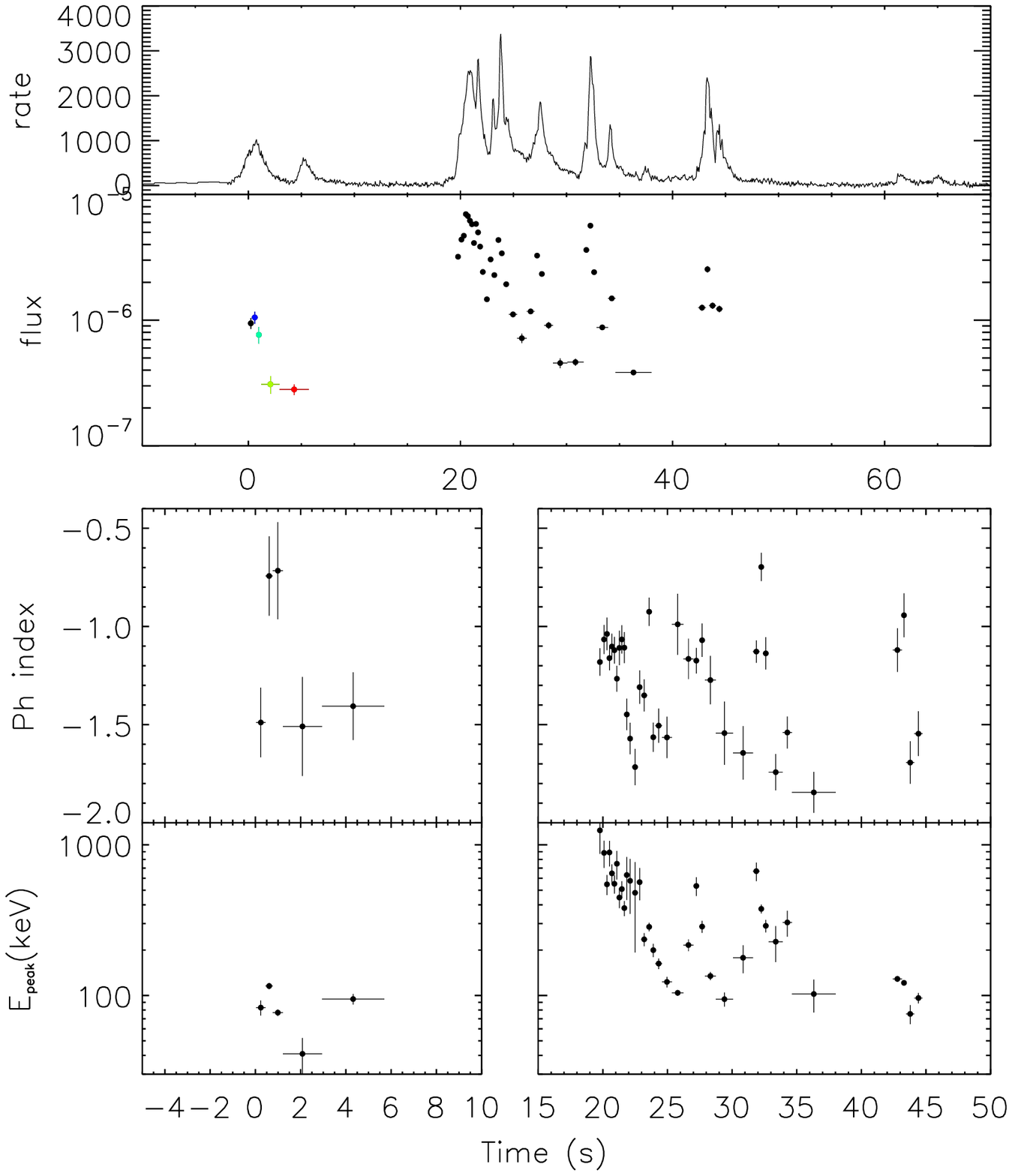}} 
\caption{Trigger \#3241. Colour code and description as in Fig. \ref{2156}--a} 
\label{3241fg1} 
\end{figure} 
\begin{figure} 
\resizebox{\hsize}{!}{\includegraphics{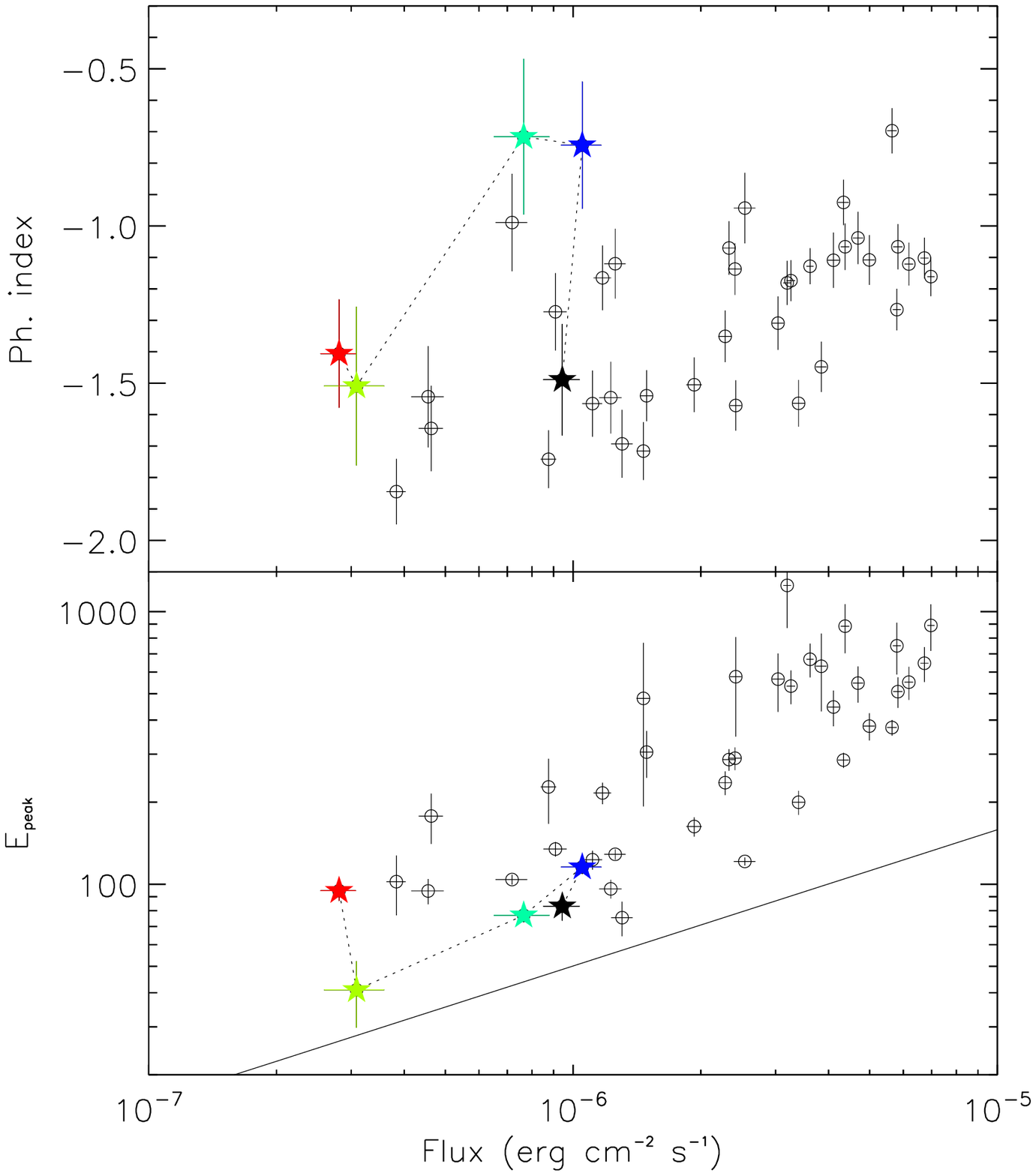}} 
\caption{Trigger \#3241. Colour code and description as in Fig. \ref{2156}--b} 
\label{3241fg2} 
\end{figure}

\begin{figure} 
\resizebox{\hsize}{!}{\includegraphics{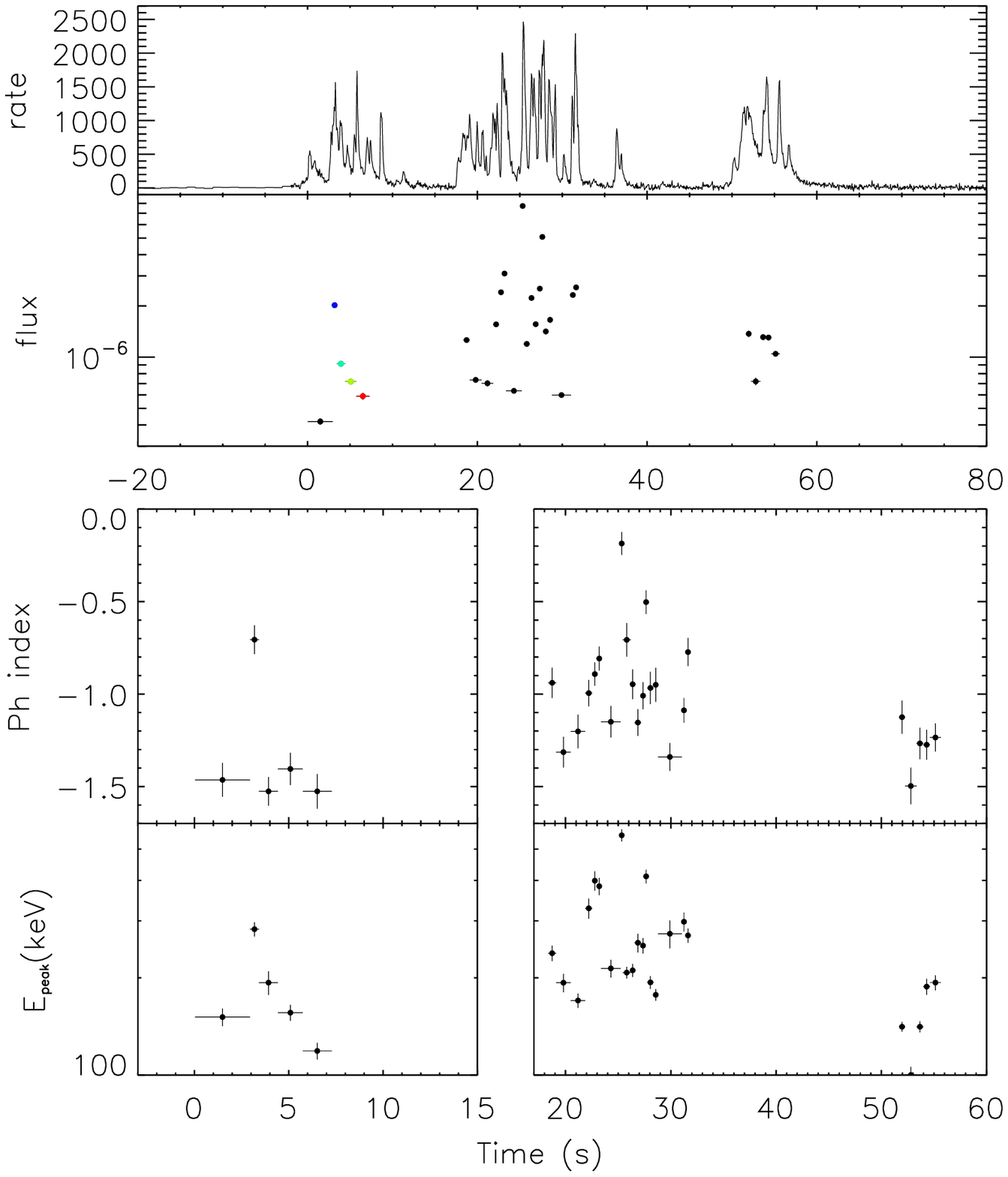}} 
\caption{Trigger \#1676. Colour code and description as in Fig. \ref{2156}--a} 
\label{1676fg1} 
\end{figure} 
\begin{figure} 
\resizebox{\hsize}{!}{\includegraphics{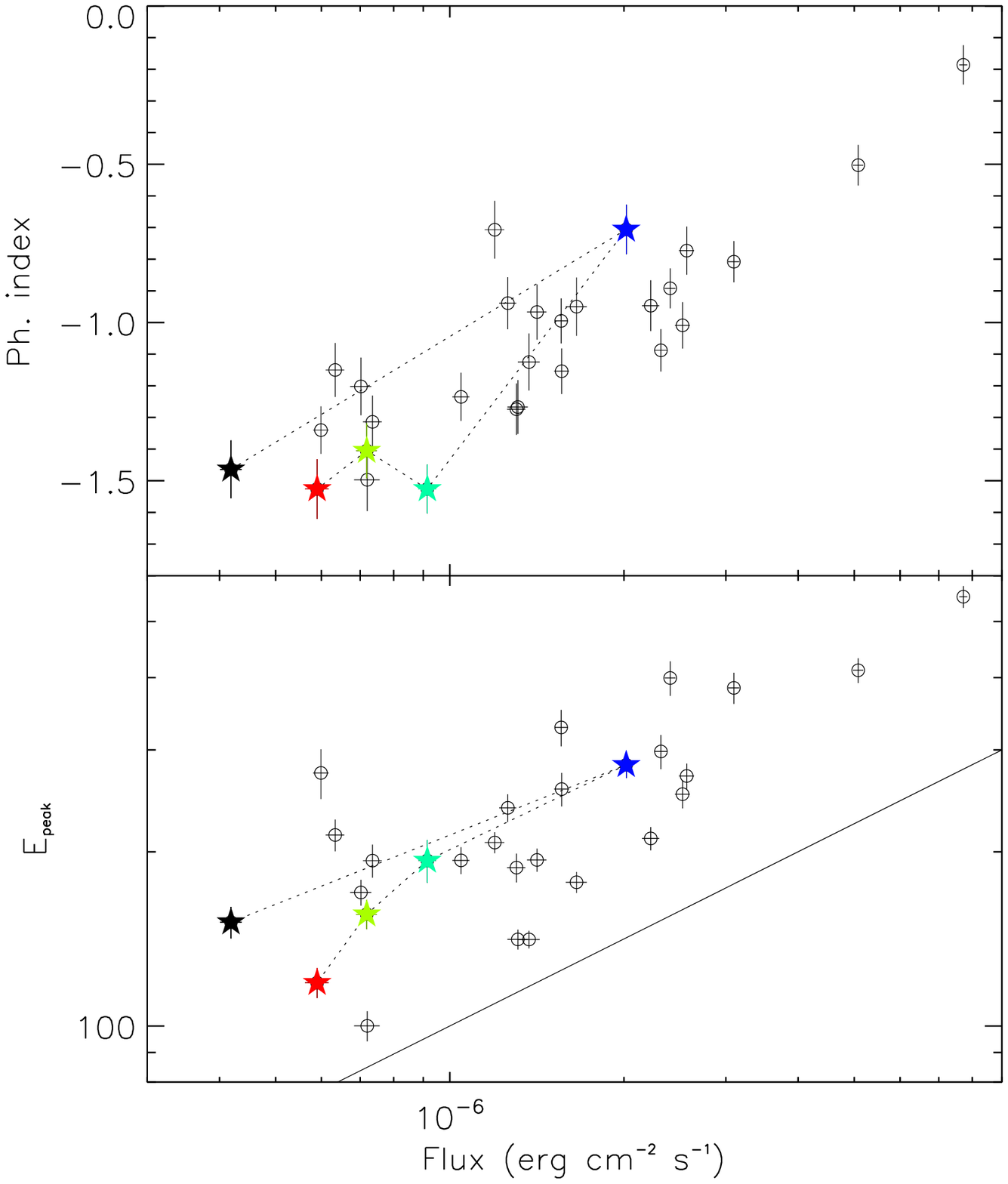}} 
\caption{Trigger \#1676. Colour code and description as in Fig. \ref{2156}--b} 
\label{1676fg2} 
\end{figure}
\clearpage
\begin{figure} 
\resizebox{\hsize}{!}{\includegraphics{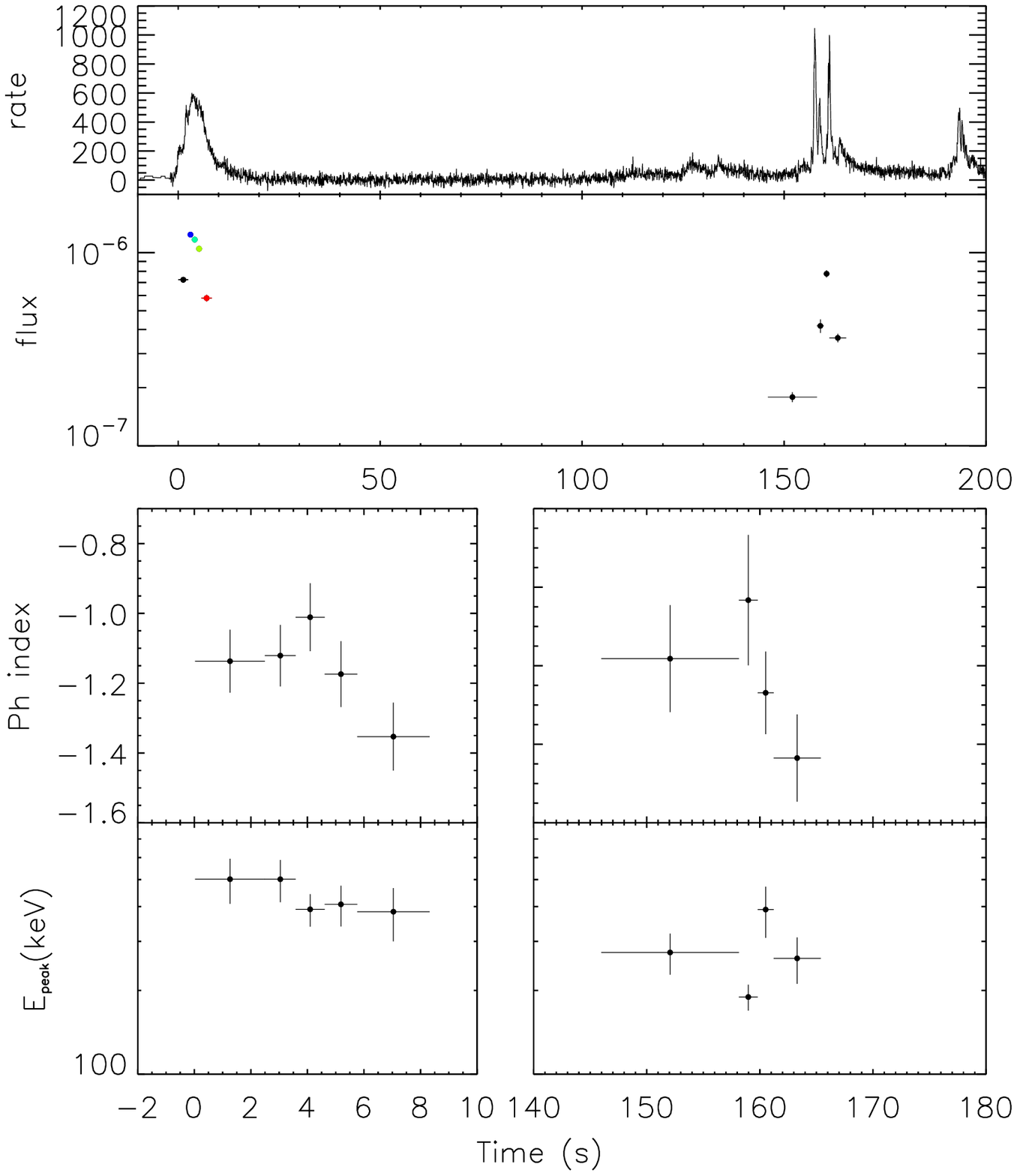}} 
\caption{Trigger \#3663. Colour code and description as in Fig. \ref{2156}--a} 
\label{3663fg1} 
\end{figure} 
\begin{figure} 
\resizebox{\hsize}{!}{\includegraphics{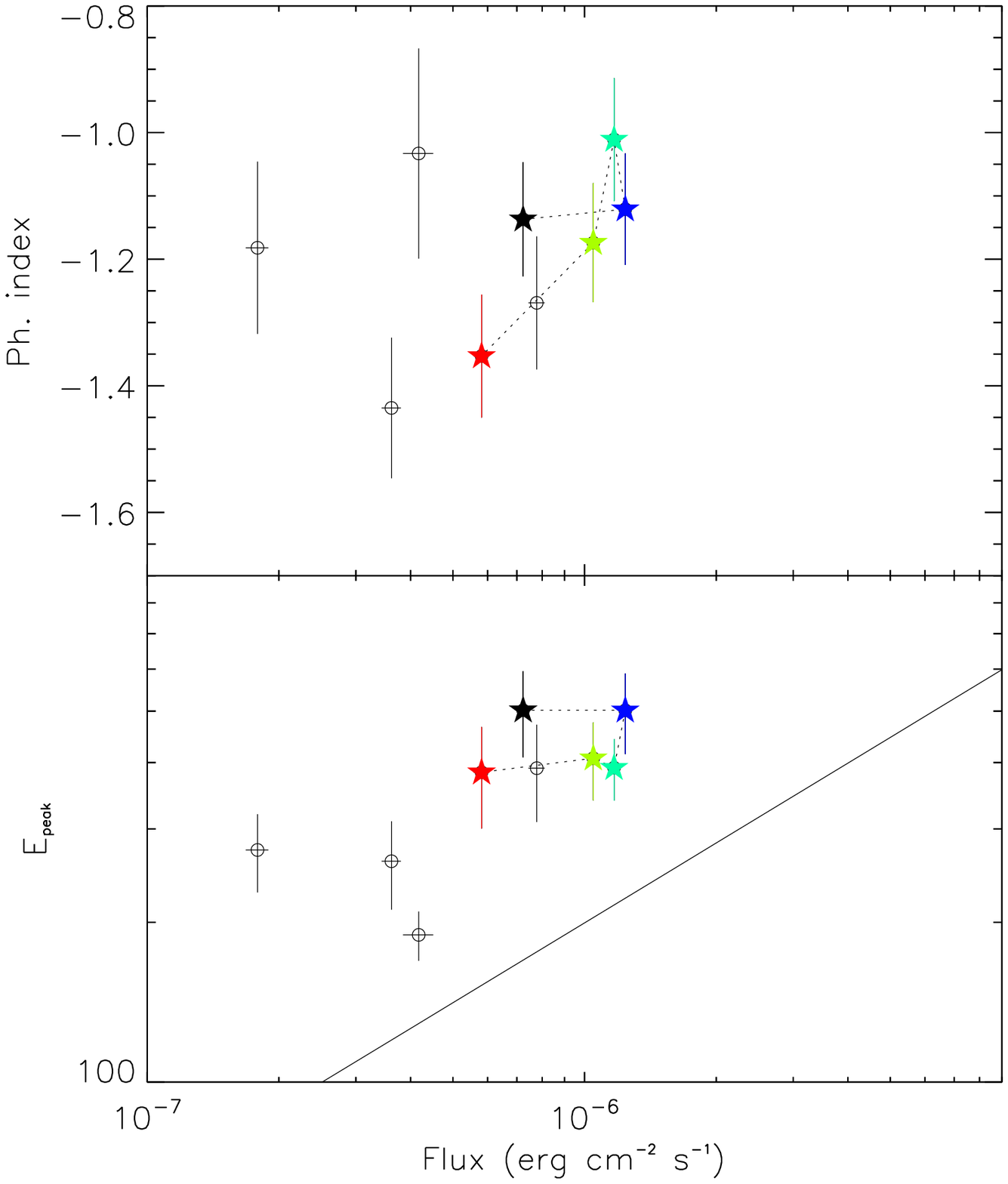}} 
\caption{Trigger \#3663. Colour code and description as in Fig. \ref{2156}--b} 
\label{3663fg2} 
\end{figure}

\begin{figure} 
\resizebox{\hsize}{!}{\includegraphics{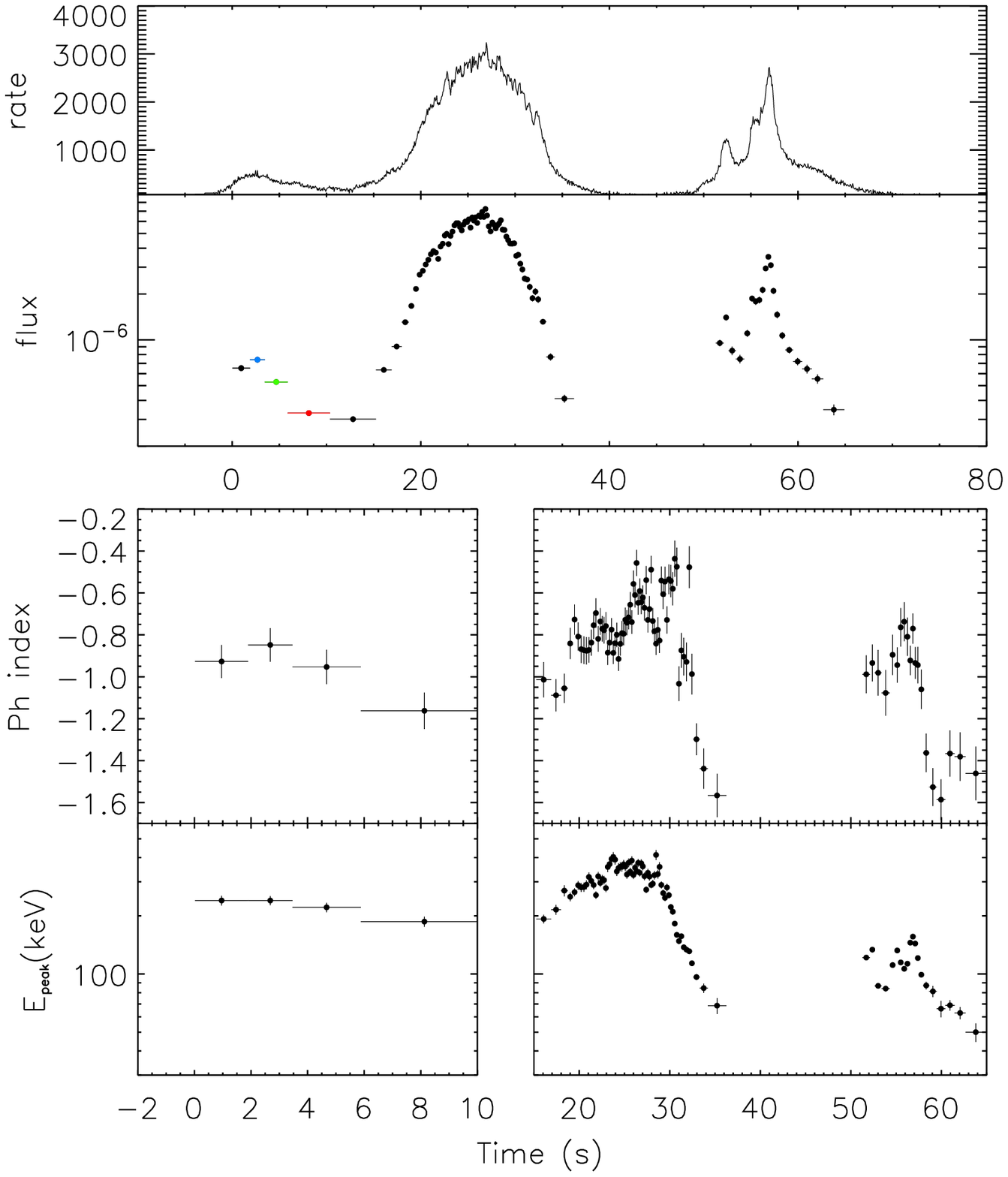}} 
\caption{Trigger \#3253. Colour code and description as in Fig. \ref{2156}--a} 
\label{3253fg1} 
\end{figure} 
\begin{figure} 
\resizebox{\hsize}{!}{\includegraphics{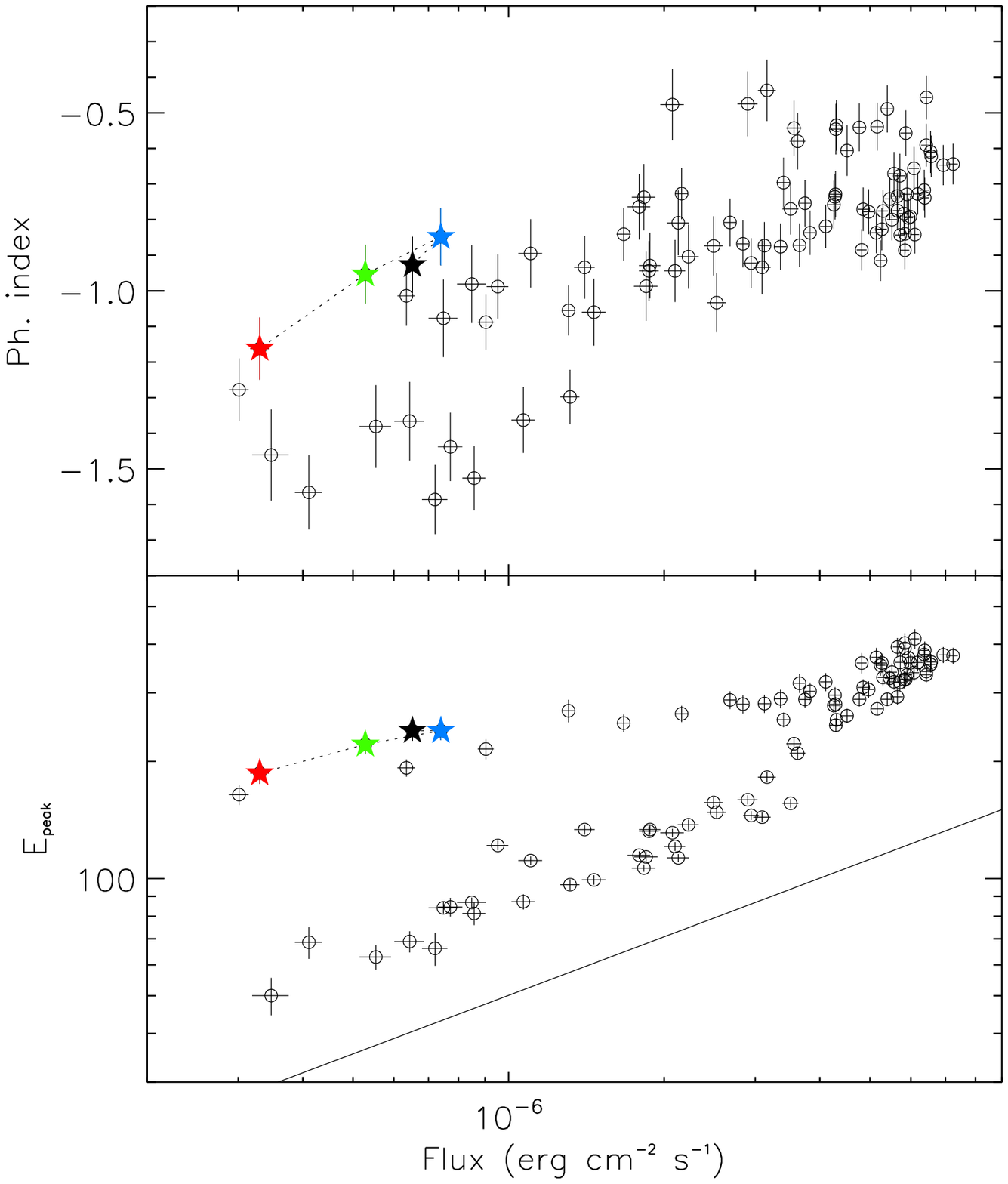}} 
\caption{Trigger \#3253. Colour code and description as in Fig. \ref{2156}--b} 
\label{3253fg2} 
\end{figure}

\begin{figure} 
\resizebox{\hsize}{!}{\includegraphics{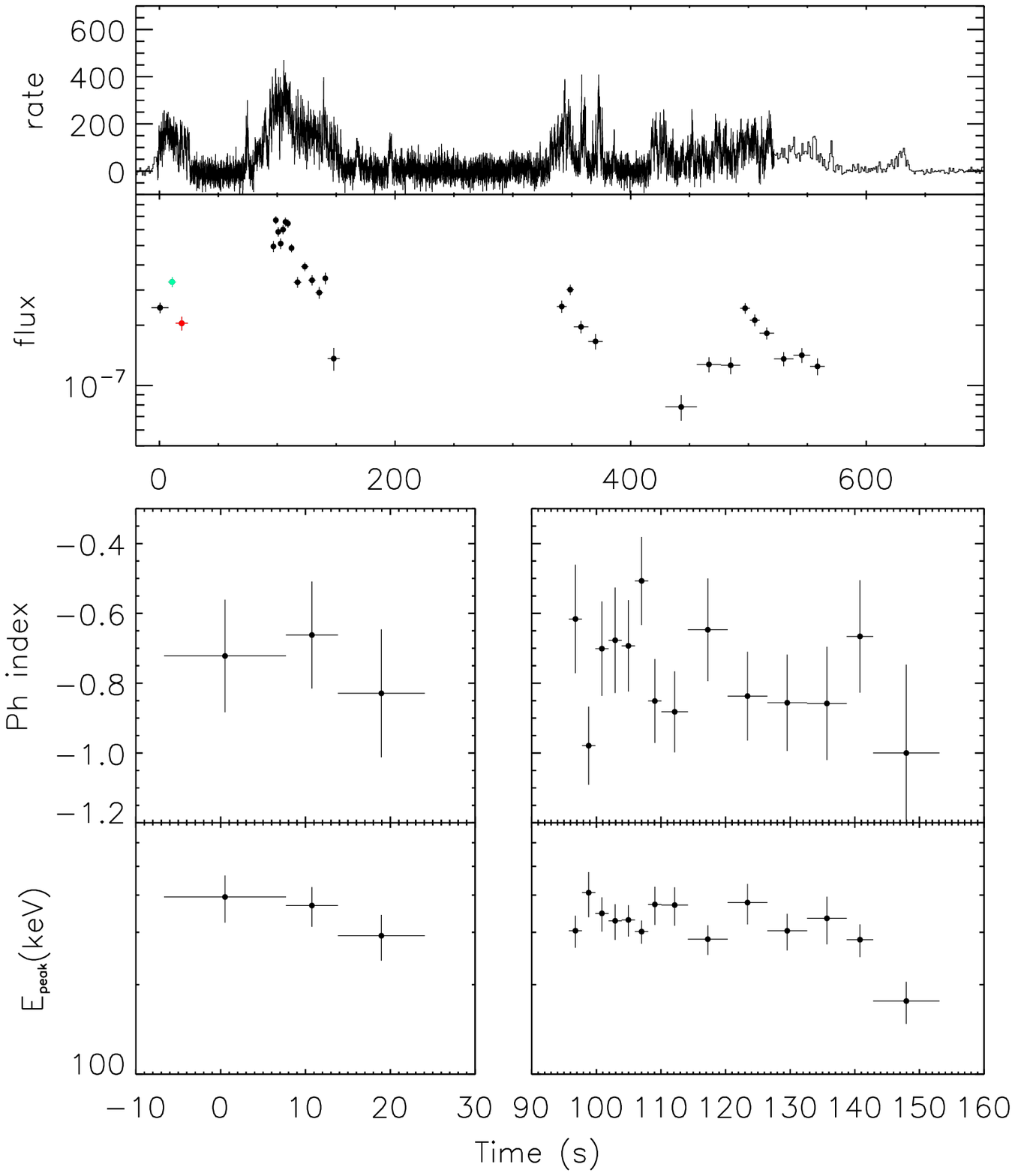}} 
\caption{Trigger \#6454. Colour code and description as in Fig. \ref{2156}--a} 
\label{6454fg1} 
\end{figure} 
\begin{figure} 
\resizebox{\hsize}{!}{\includegraphics{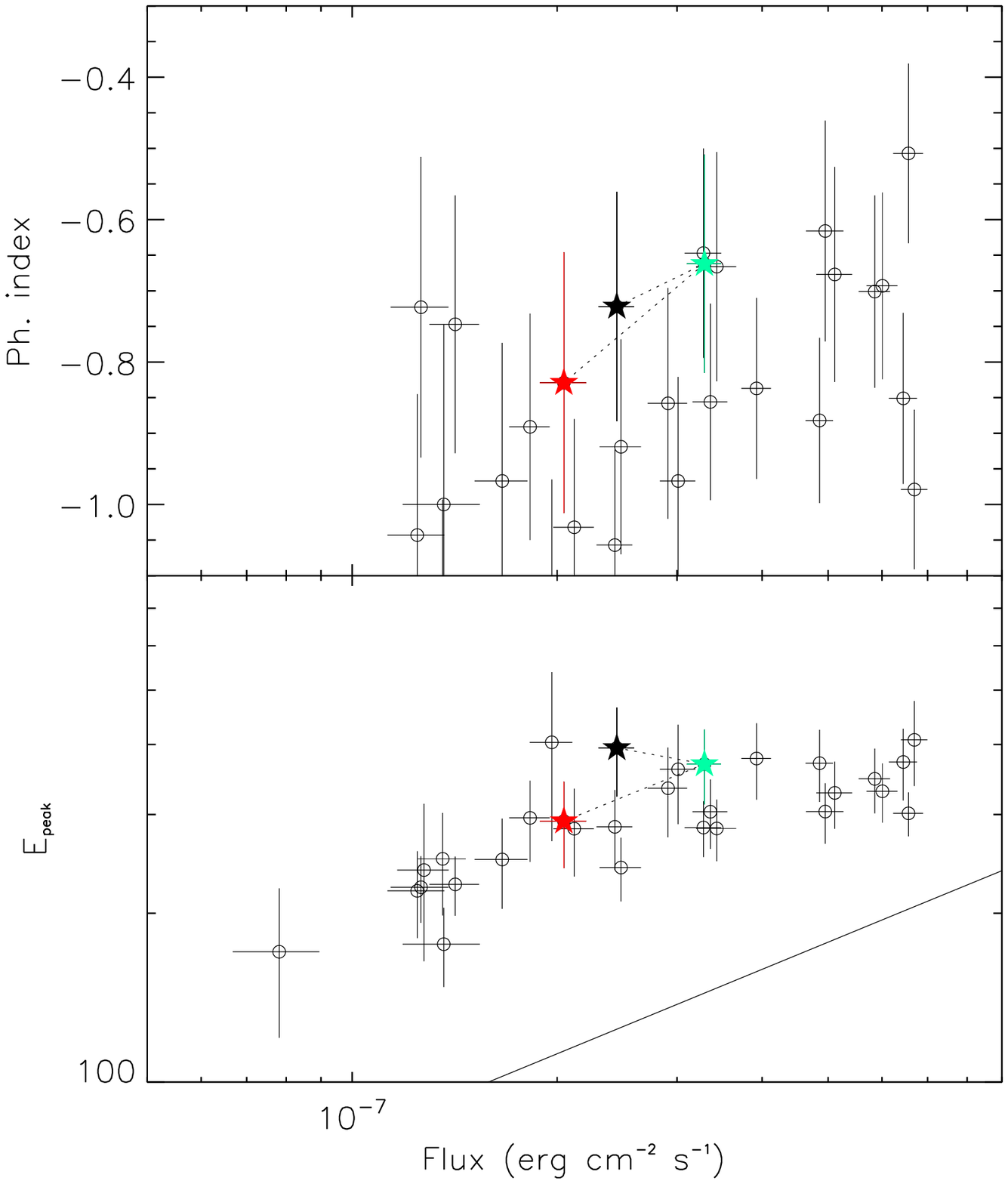}} 
\caption{Trigger \#6454. Colour code and description as in Fig. \ref{2156}--b} 
\label{6454fg2} 
\end{figure}

\begin{figure} 
\resizebox{\hsize}{!}{\includegraphics{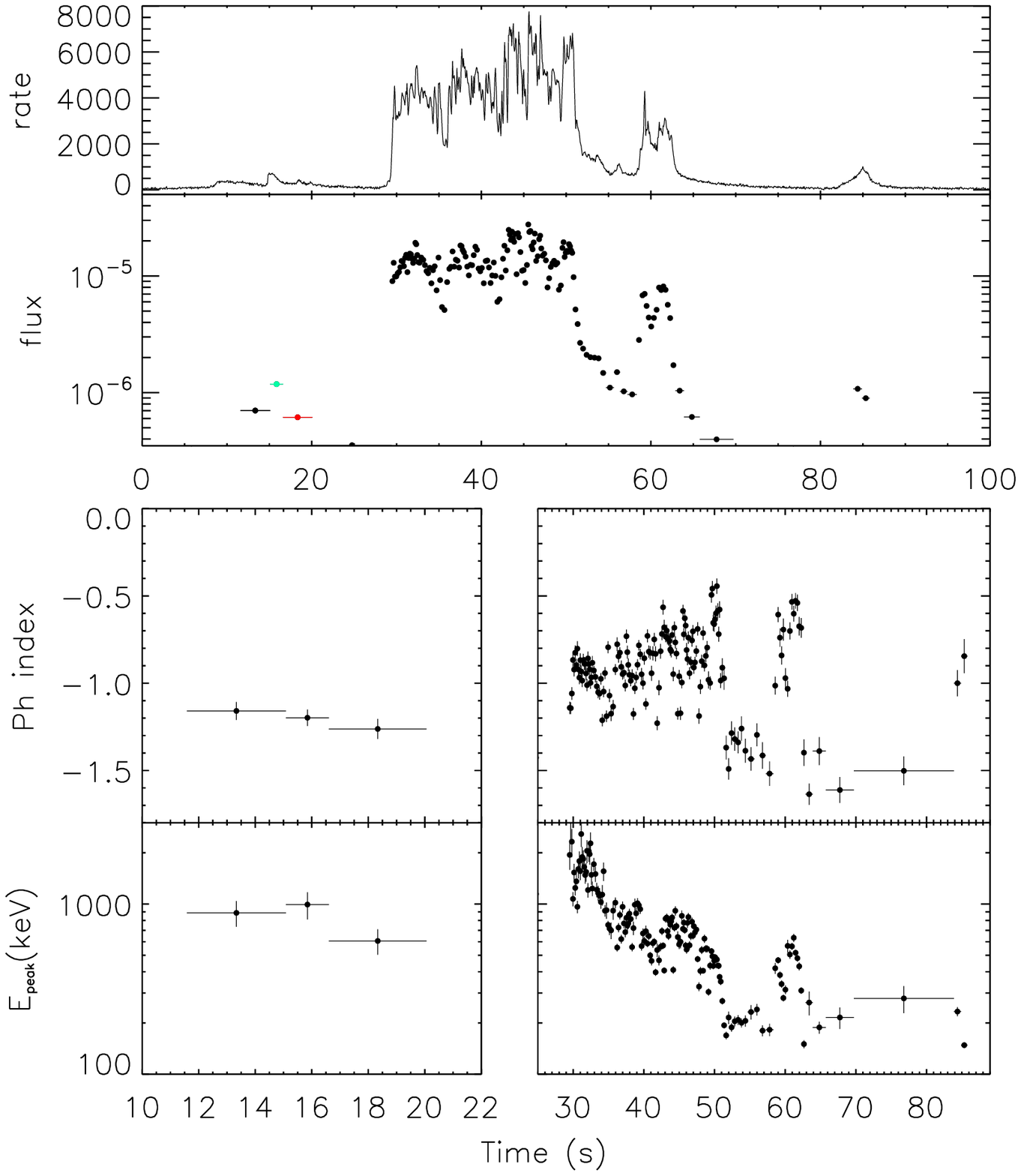}} 
\caption{Trigger \#3057. Colour code and description as in Fig. \ref{2156}--a} 
\label{3057fg1} 
\end{figure} 
\begin{figure} 
\resizebox{\hsize}{!}{\includegraphics{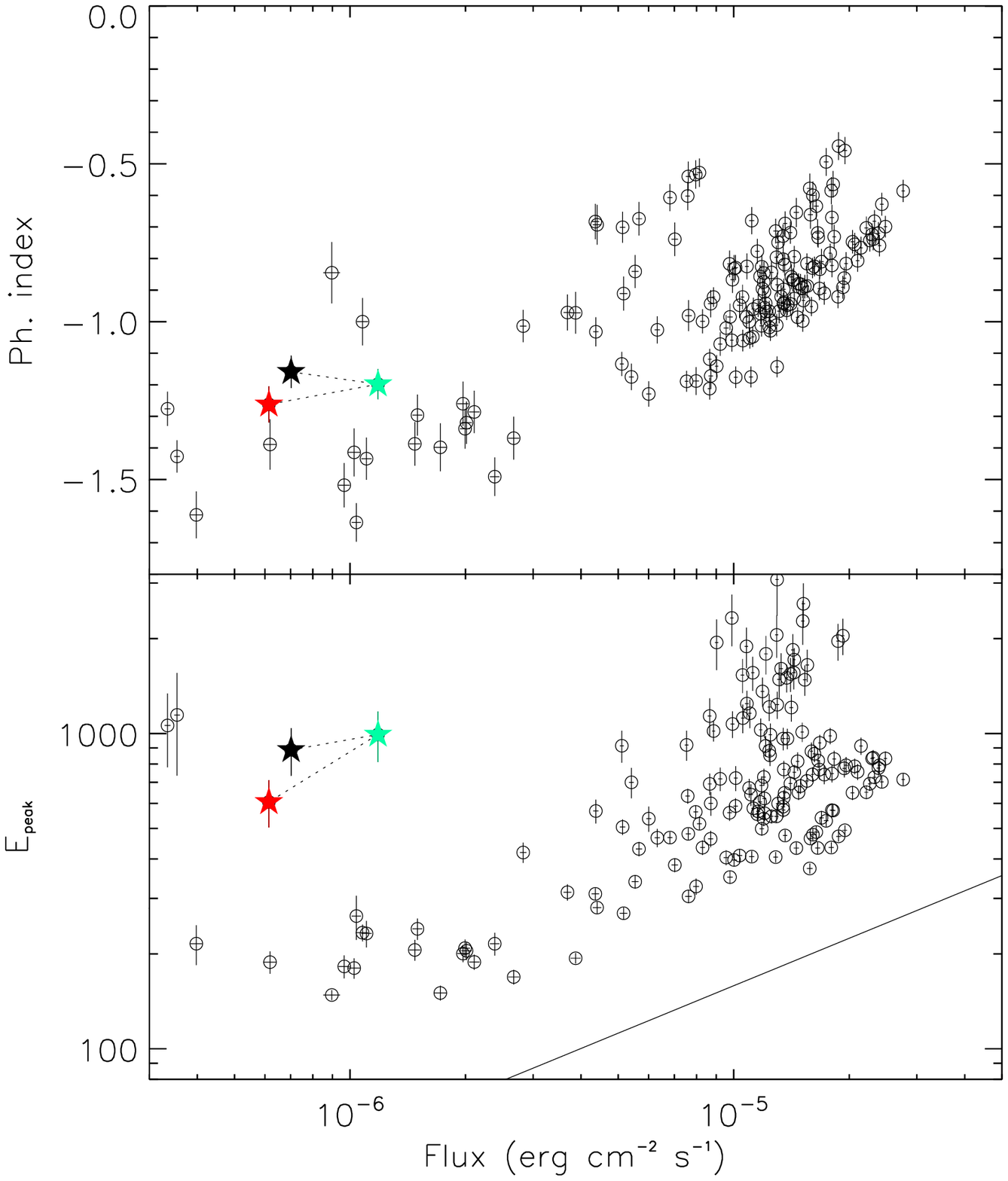}} 
\caption{Trigger \#3057. Colour code and description as in Fig. \ref{2156}--b} 
\label{3057fg2} 
\end{figure}

\begin{figure} 
\resizebox{\hsize}{!}{\includegraphics{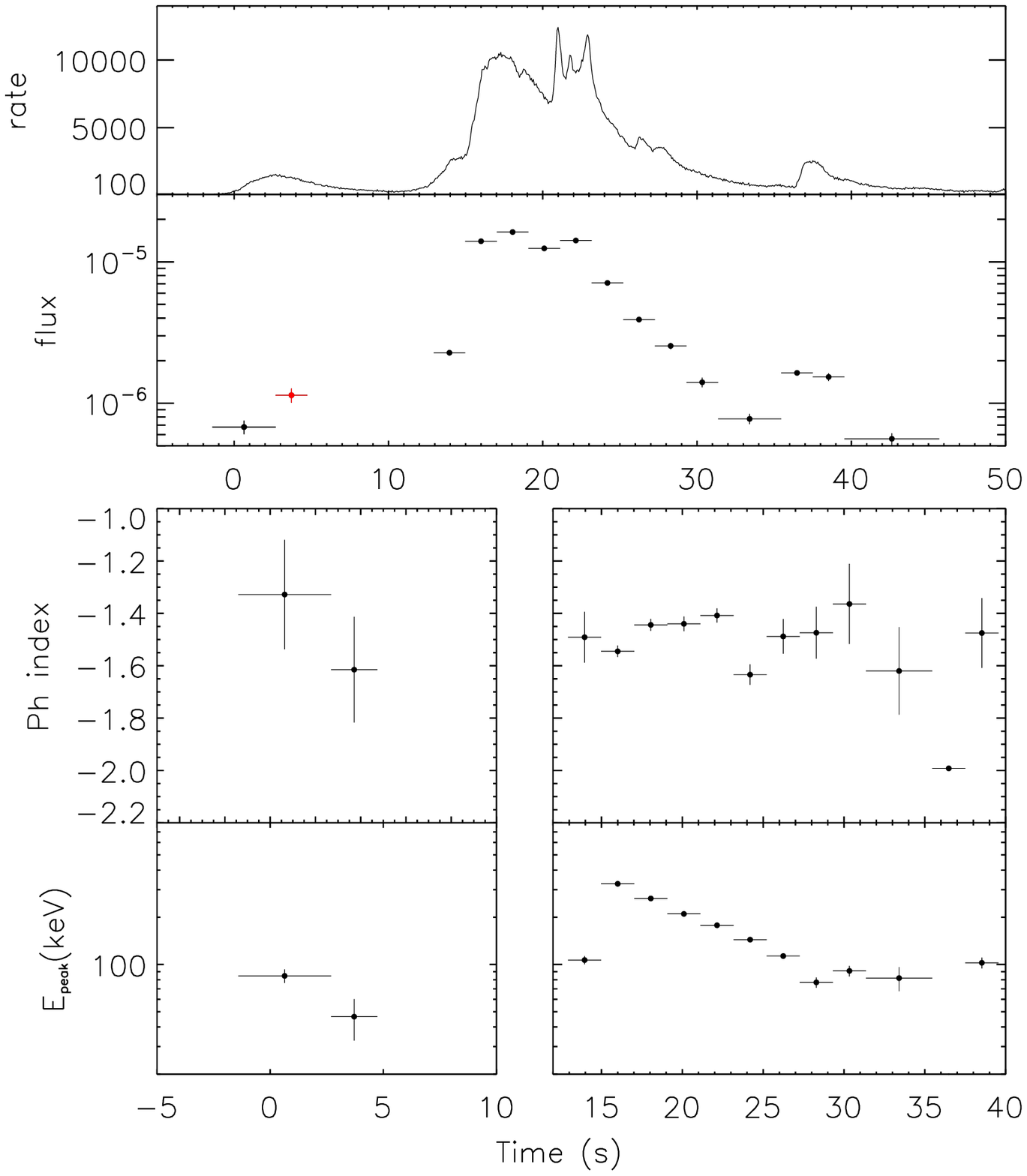}} 
\caption{Trigger \#4368. Colour code and description as in Fig. \ref{2156}--a} 
\label{4368fg1} 
\end{figure} 
\begin{figure} 
\resizebox{\hsize}{!}{\includegraphics{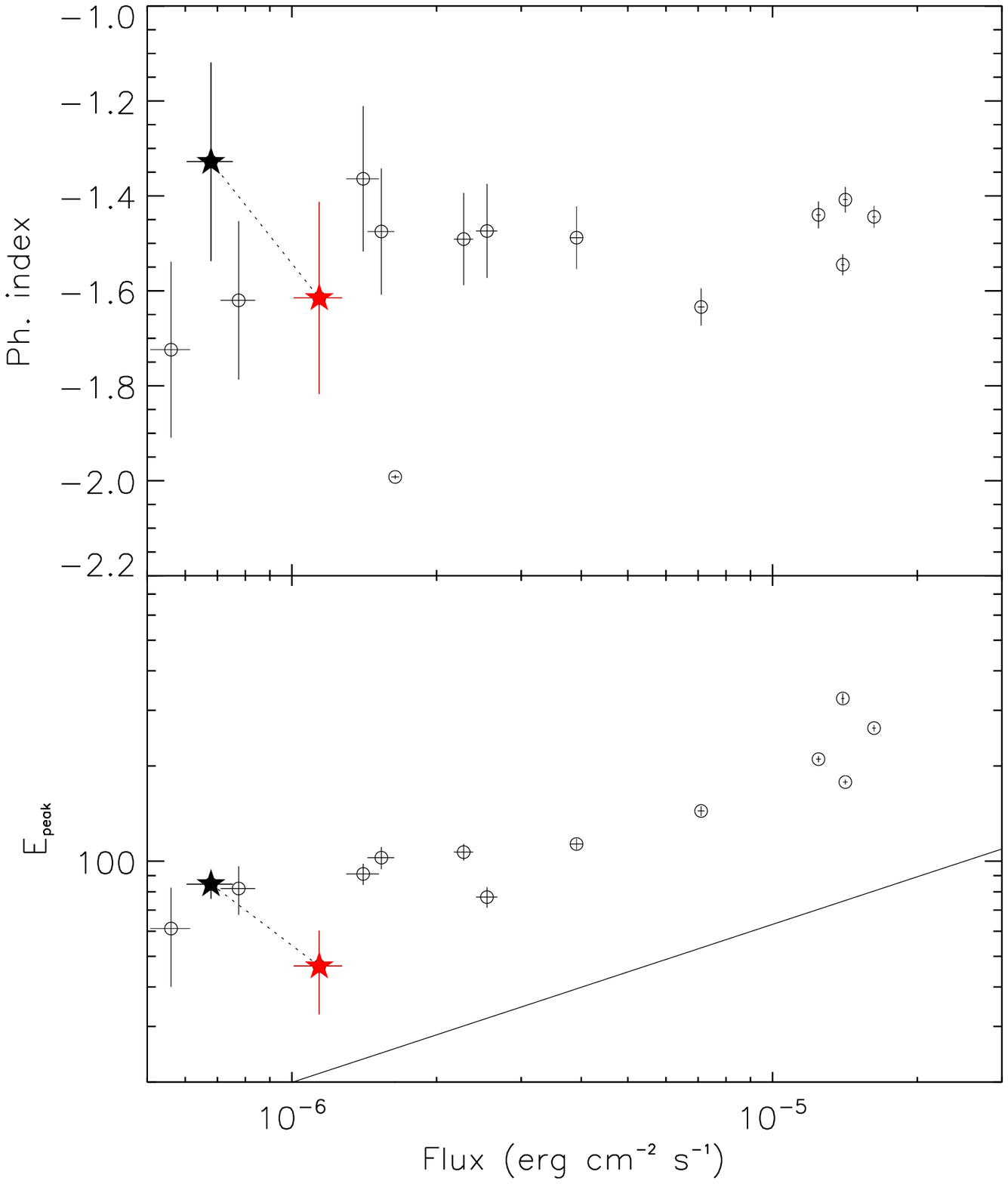}} 
\caption{Trigger \#4368. Colour code and description as in Fig. \ref{2156}--b} 
\label{4368fg2} 
\end{figure} 

\begin{figure} 
\resizebox{\hsize}{!}{\includegraphics{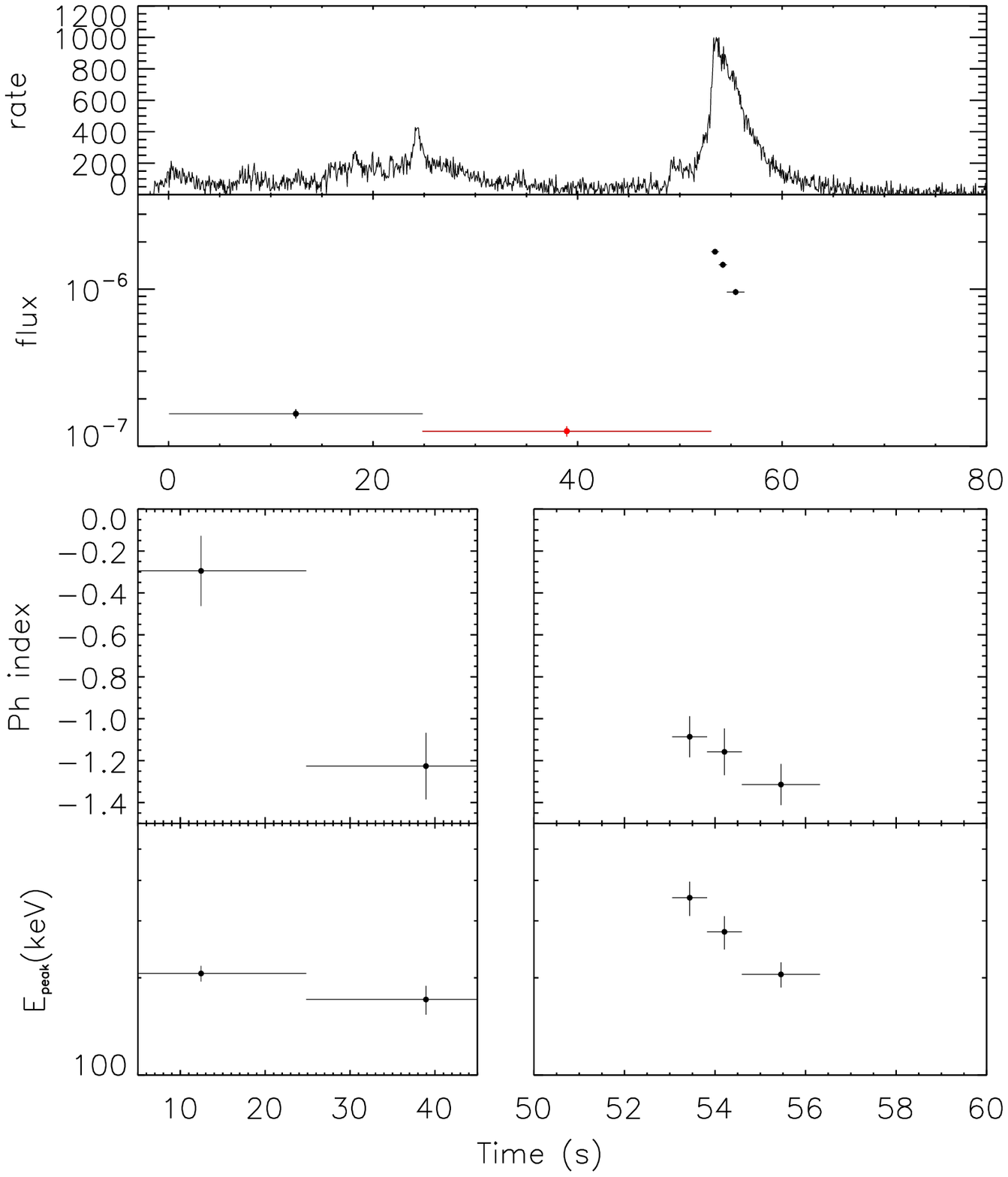}} 
\caption{Trigger \#2700. Colour code and description as in Fig. \ref{2156}--a} 
\label{2700fg1} 
\end{figure} 
\begin{figure} 
\resizebox{\hsize}{!}{\includegraphics{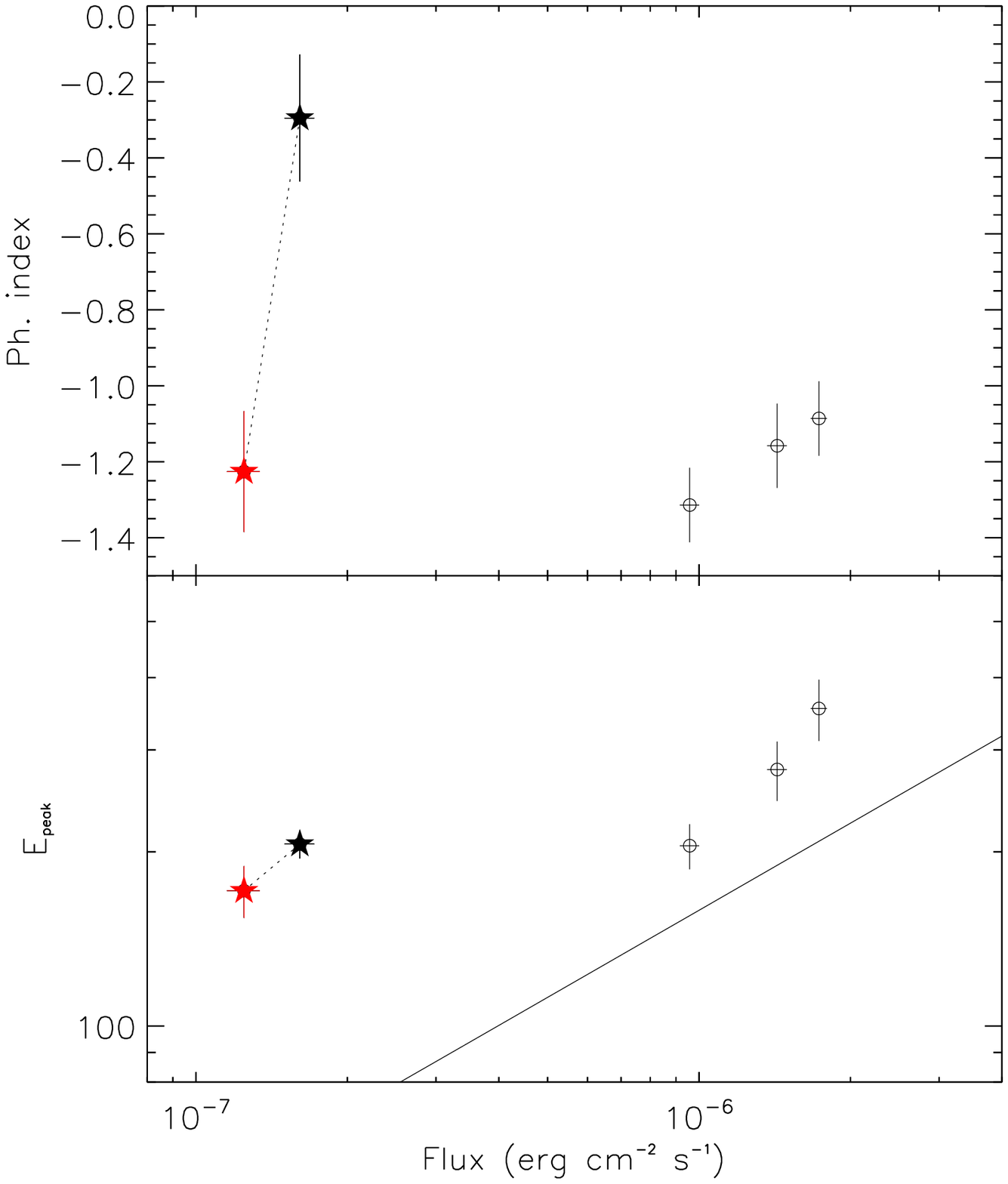}} 
\caption{Trigger \#2700. Colour code and description as in Fig. \ref{2156}--b} 
\label{2700fg2} 
\end{figure}

\begin{figure} 
\resizebox{\hsize}{!}{\includegraphics{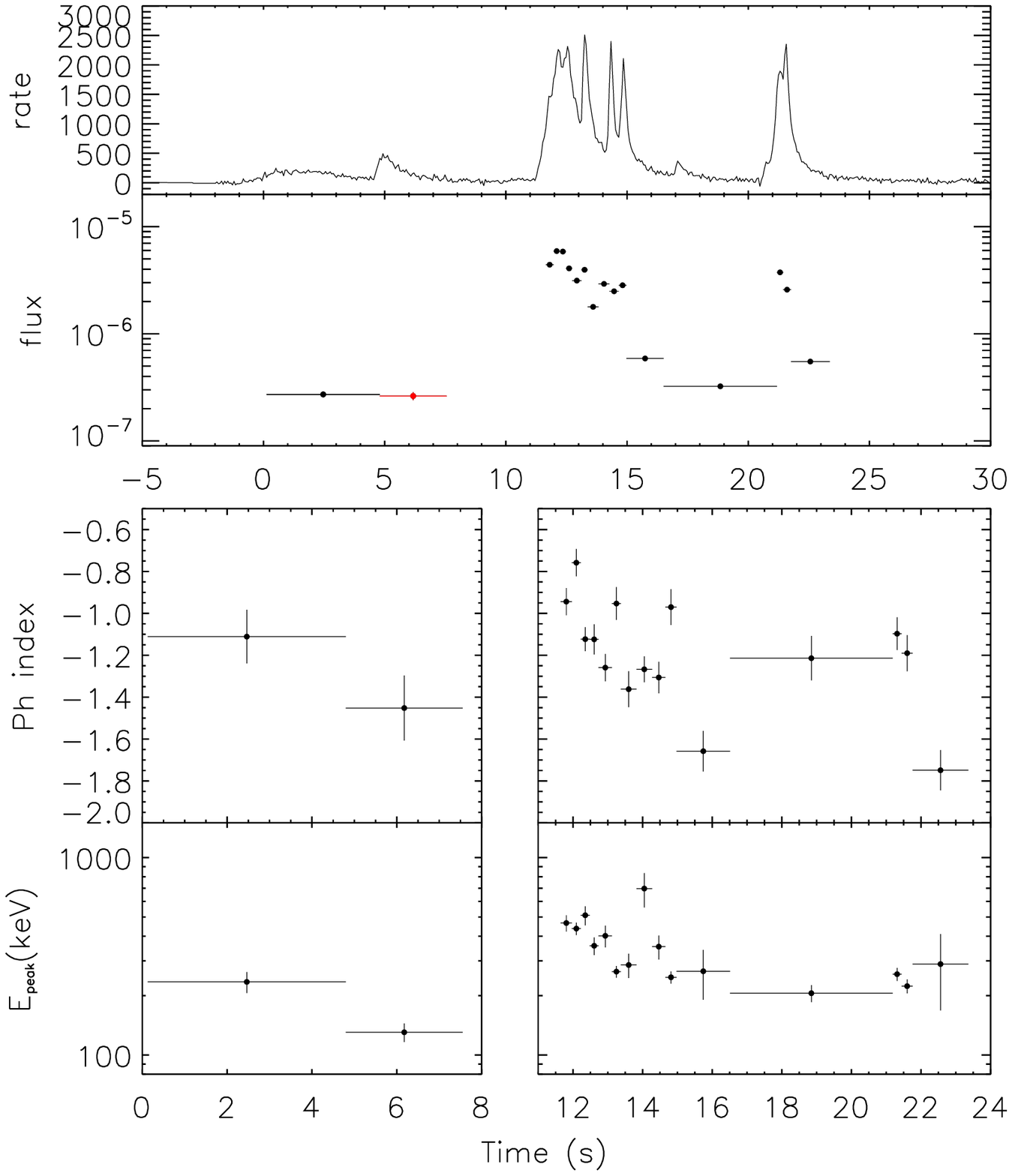}} 
\caption{Trigger \#1157. Colour code and description as in Fig. \ref{2156}--a} 
\label{1157fg1} 
\end{figure} 
\begin{figure} 
\resizebox{\hsize}{!}{\includegraphics{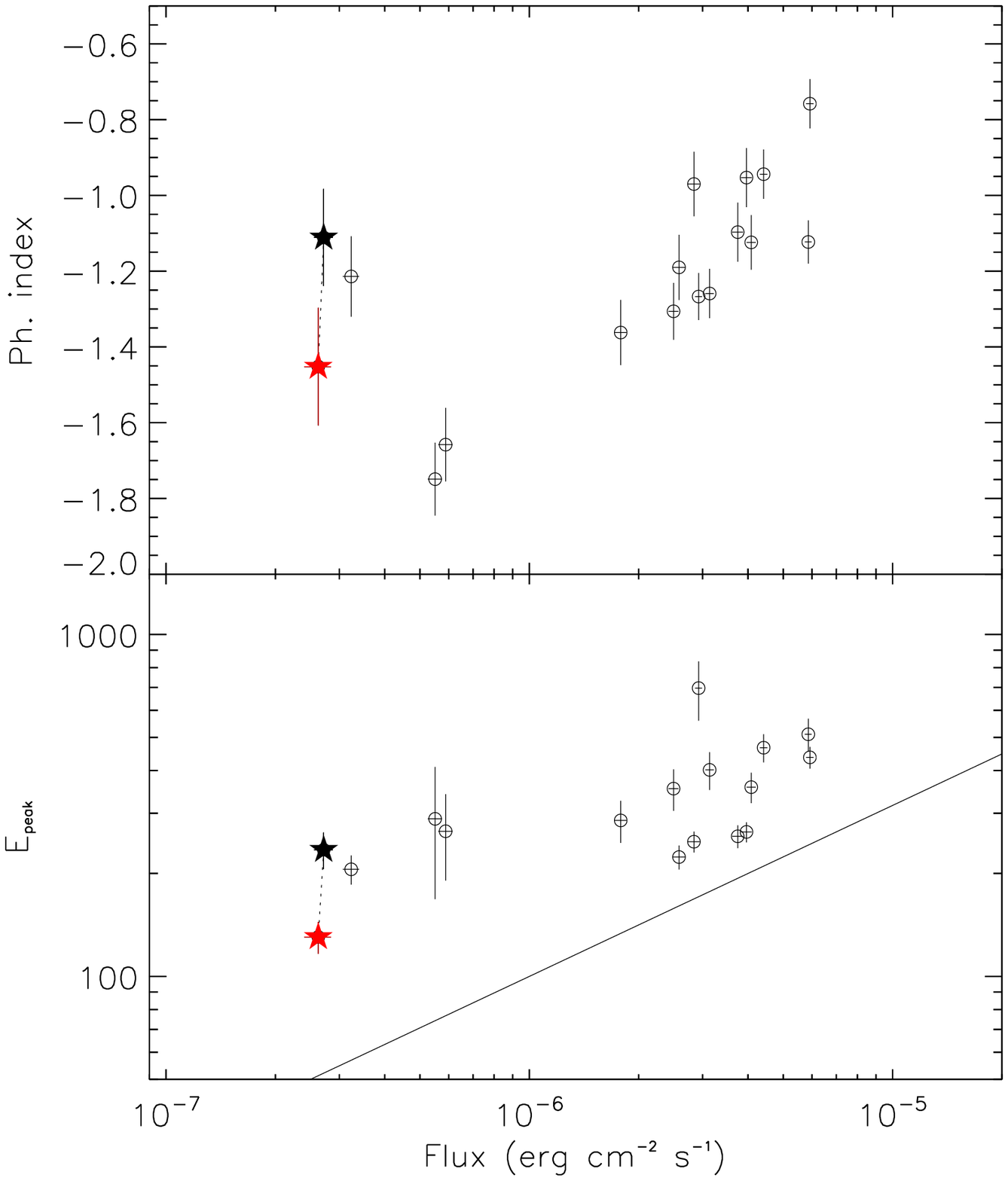}} 
\caption{Trigger \#1157. Colour code and description as in Fig. \ref{2156}--b} 
\label{1157fg2} 
\end{figure}

\begin{figure} 
\resizebox{\hsize}{!}{\includegraphics{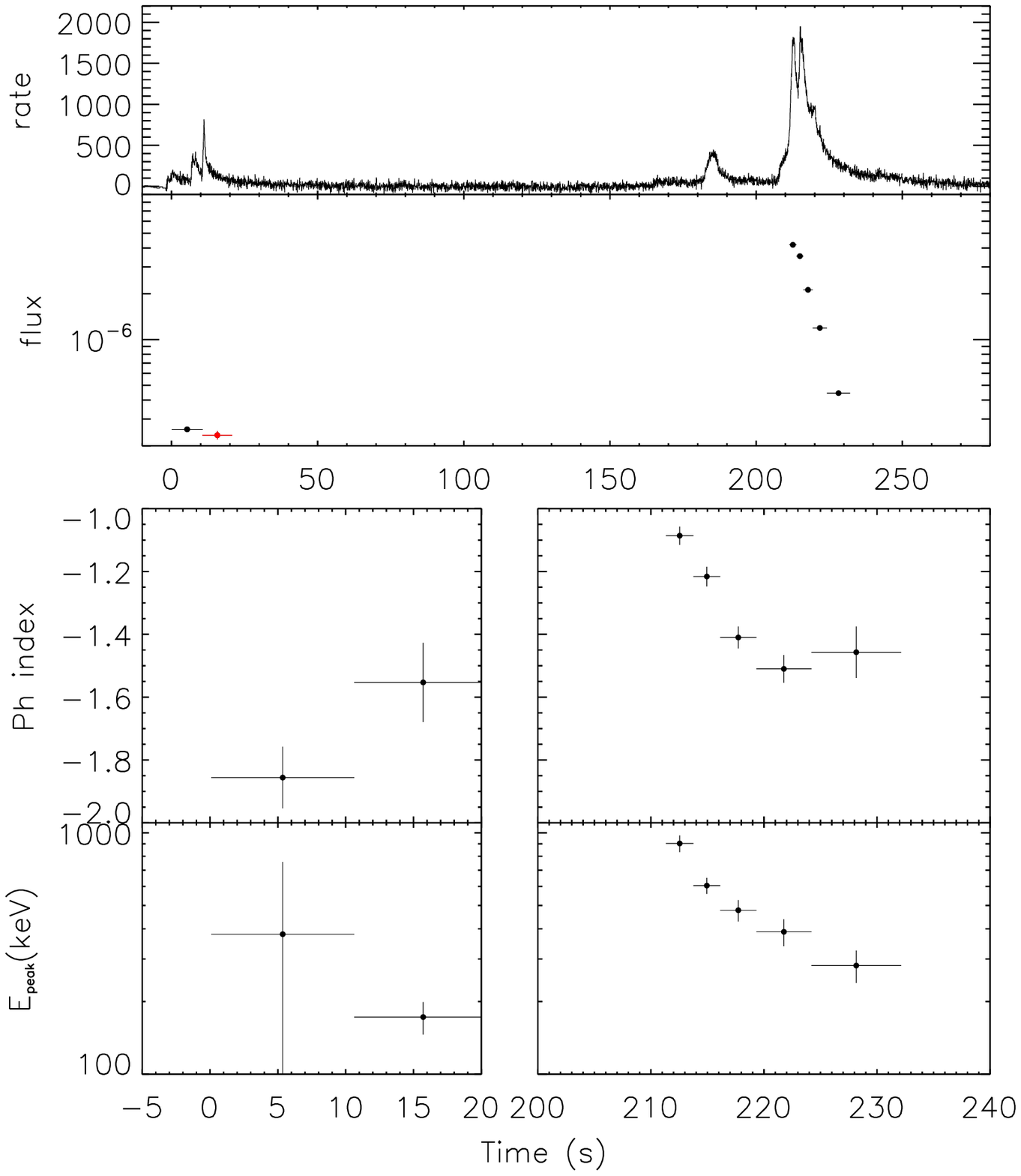}} 
\caption{Trigger \#6629. Colour code and description as in Fig. \ref{2156}--a} 
\label{6629fg1} 
\end{figure} 
\begin{figure} 
\resizebox{\hsize}{!}{\includegraphics{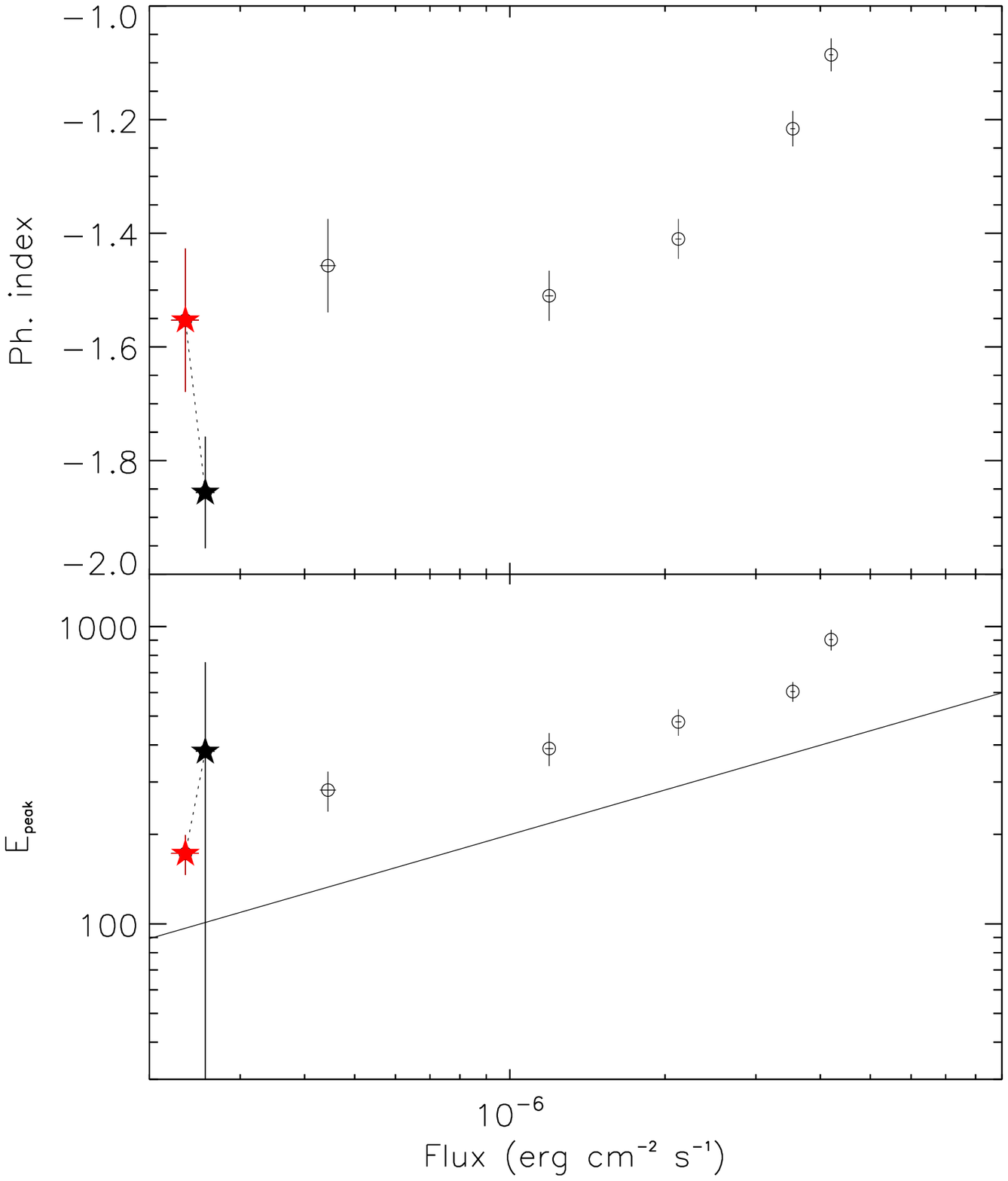}} 
\caption{Trigger \#6629. Colour code and description as in Fig. \ref{2156}--b} 
\label{6629fg2} 
\end{figure}
\clearpage
\begin{figure} 
\resizebox{\hsize}{!}{\includegraphics{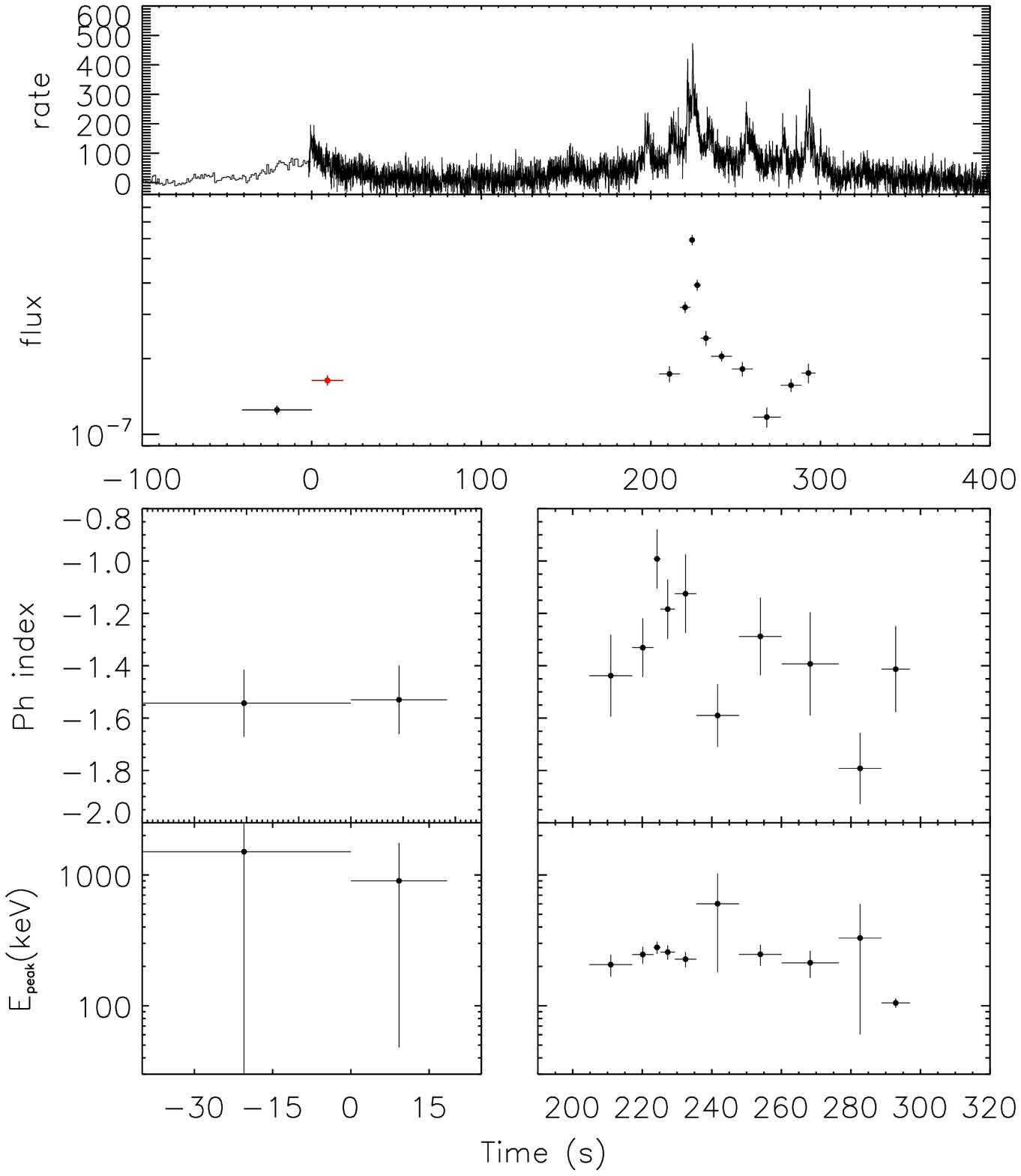}} 
\caption{Trigger \#3448. Colour code and description as in Fig. \ref{2156}--a} 
\label{3448fg1} 
\end{figure} 
\begin{figure} 
\resizebox{\hsize}{!}{\includegraphics{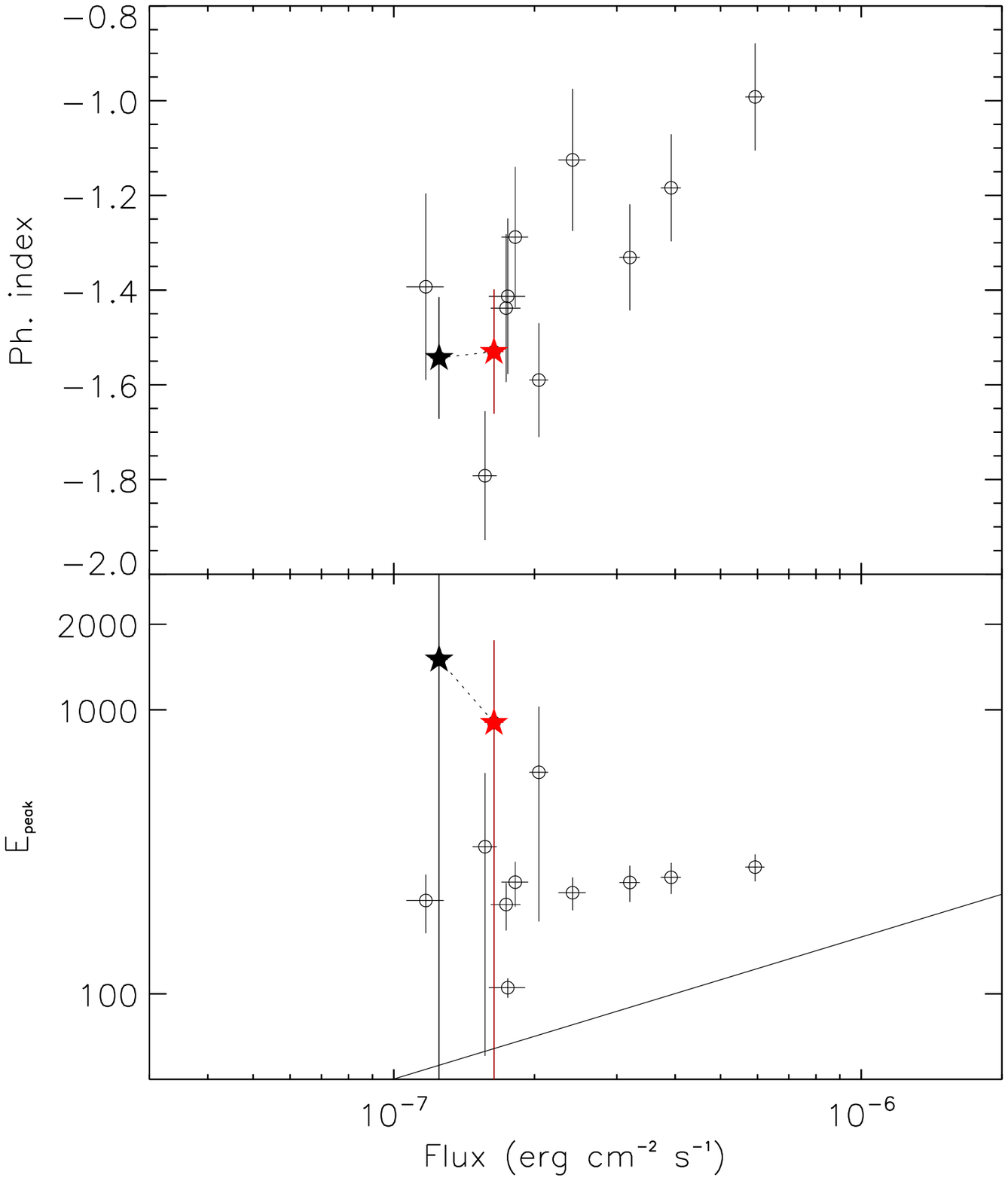}} 
\caption{Trigger \#3448. Colour code and description as in Fig. \ref{2156}--b} 
\label{3448fg2} 
\end{figure}

\begin{figure} 
\resizebox{\hsize}{!}{\includegraphics{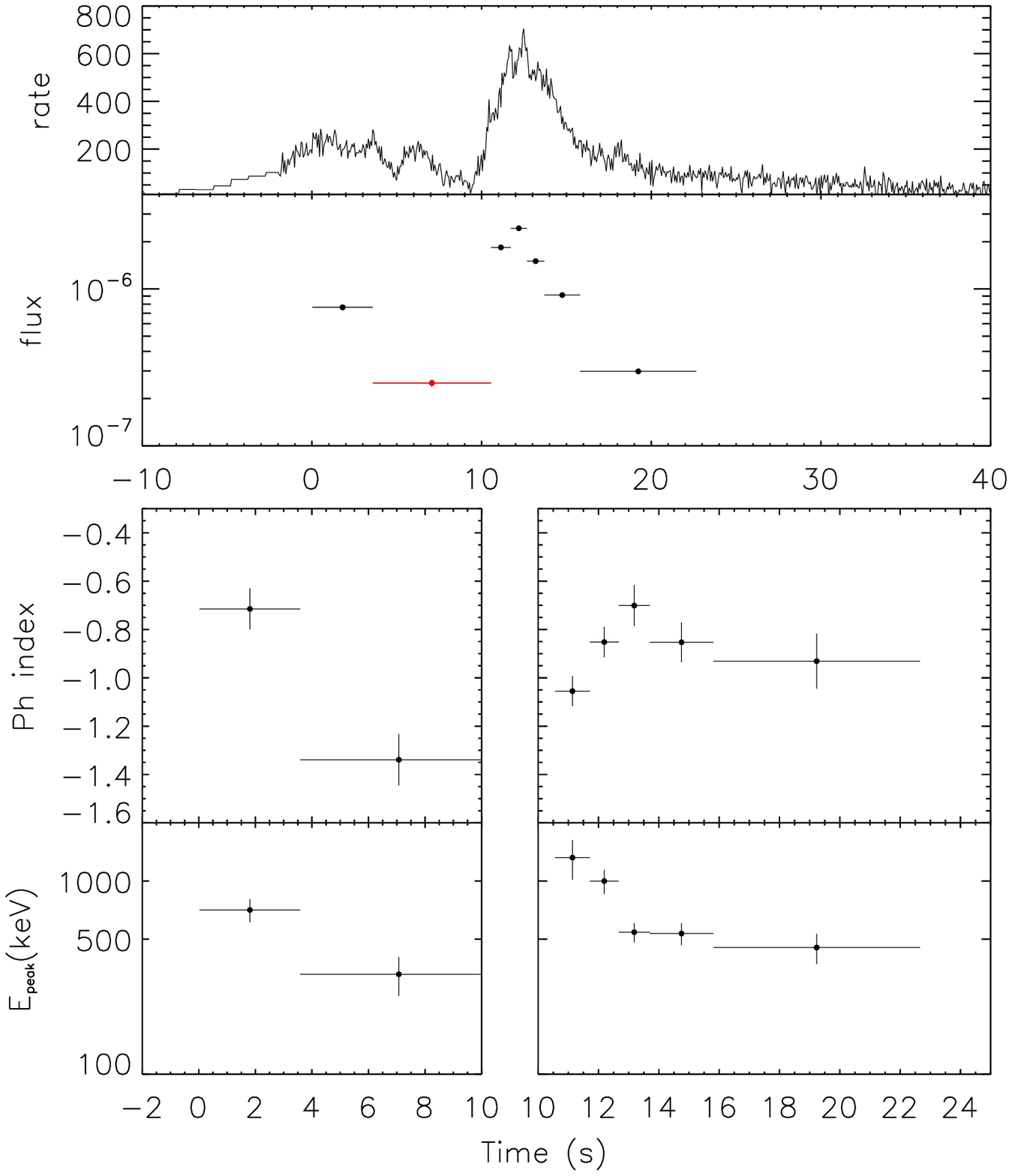}} 
\caption{Trigger \#3301. Colour code and description as in Fig. \ref{2156}--a} 
\label{3301fg1} 
\end{figure} 
\begin{figure} 
\resizebox{\hsize}{!}{\includegraphics{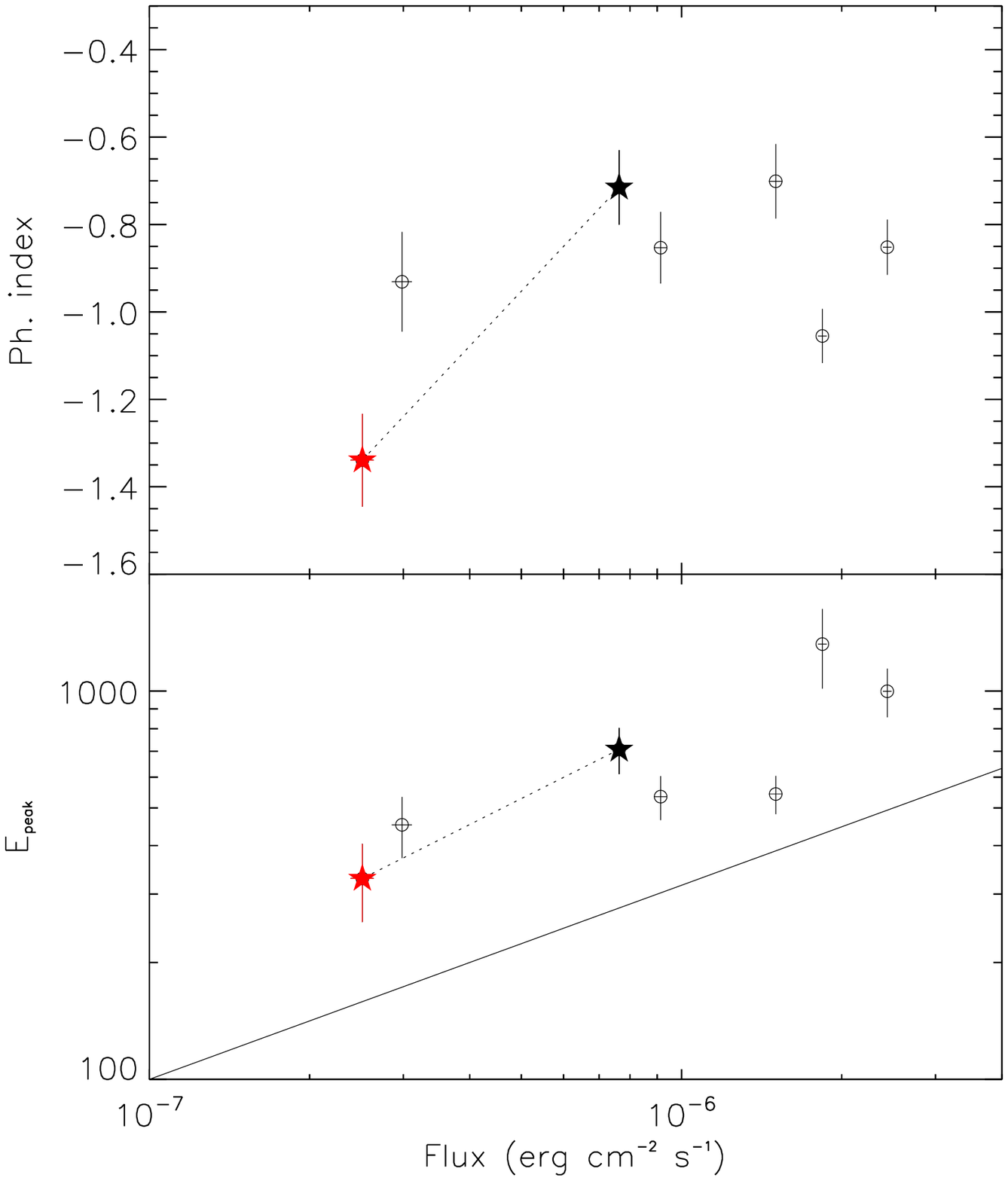}} 
\caption{Trigger \#3301. Colour code and description as in Fig. \ref{2156}--b} 
\label{3301fg2} 
\end{figure}

\begin{figure} 
\resizebox{\hsize}{!}{\includegraphics{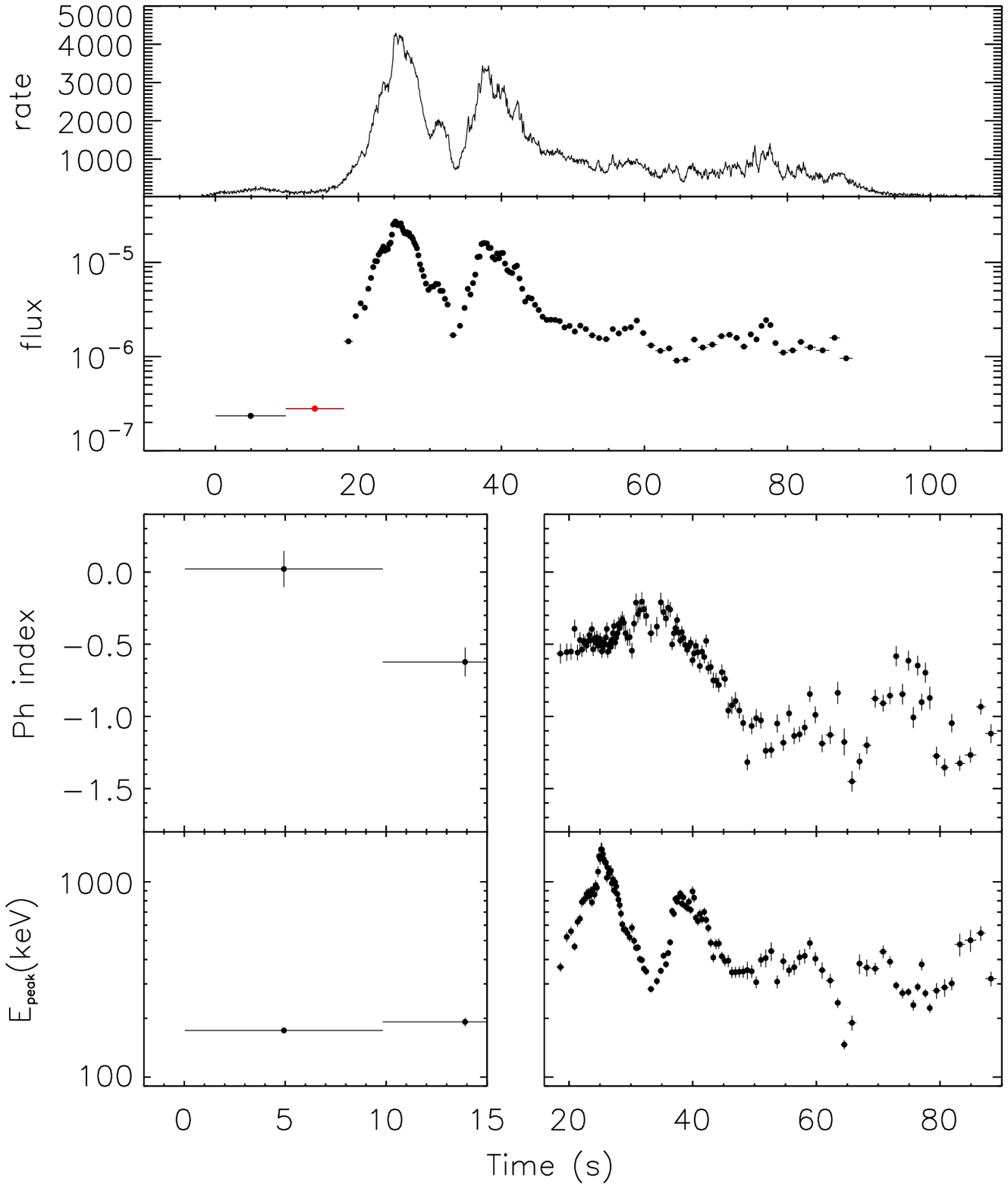}} 
\caption{Trigger \#7343. Colour code and description as in Fig. \ref{2156}--a} 
\label{7343fg1} 
\end{figure} 
\begin{figure} 
\resizebox{\hsize}{!}{\includegraphics{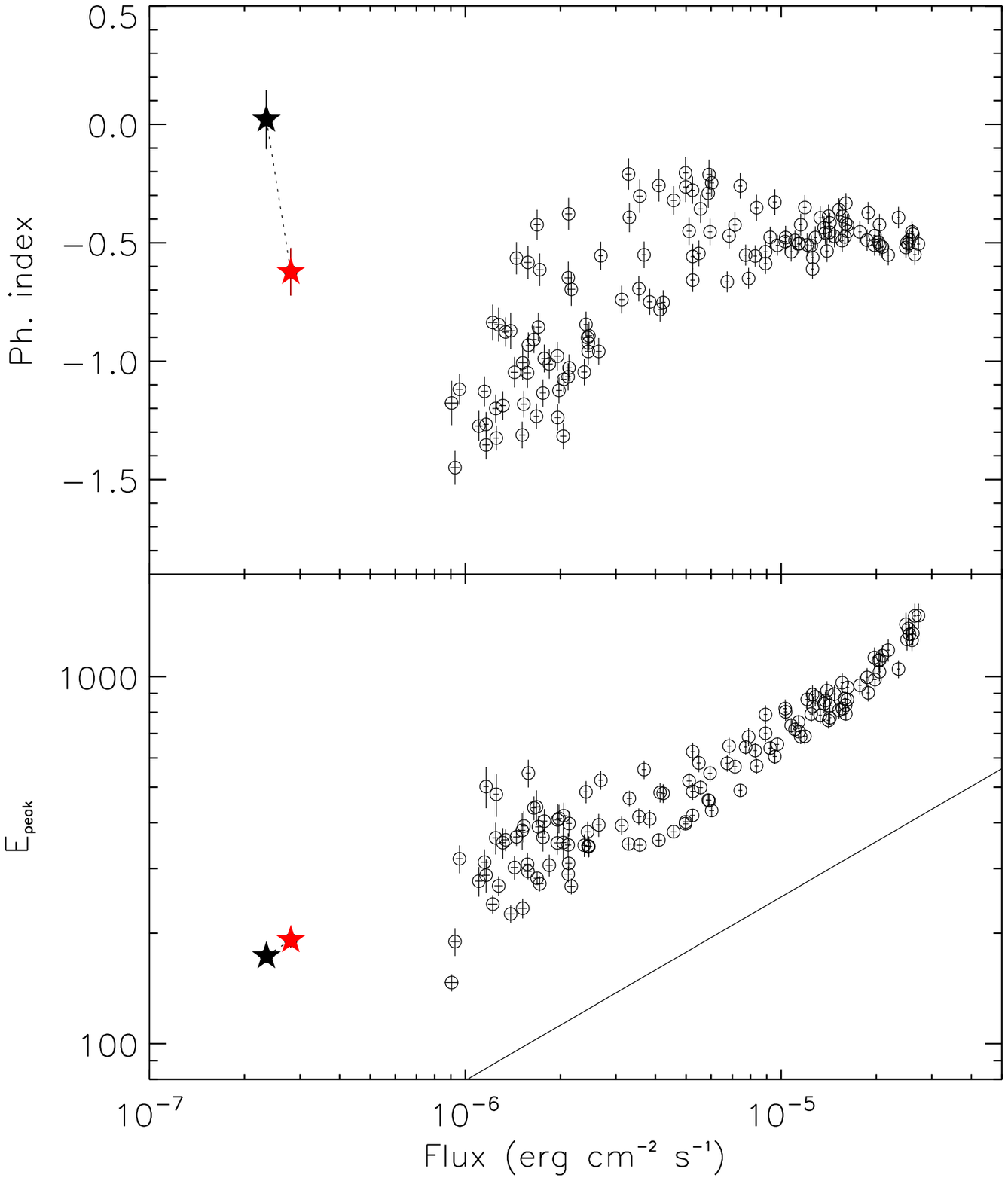}} 
\caption{Trigger \#7343. Colour code and description as in Fig. \ref{2156}--b} 
\label{7343fg2} 
\end{figure}

\end{document}